\documentclass{SciPost}
\usepackage{comment}
\hypersetup{
    colorlinks,
    linkcolor={red!50!black},
    citecolor={blue!50!black},
    urlcolor={blue!80!black}
}

\usepackage[bitstream-charter]{mathdesign}
\DeclareSymbolFont{usualmathcal}{OMS}{cmsy}{m}{n}
\DeclareSymbolFontAlphabet{\mathcal}{usualmathcal}

\fancypagestyle{SPstyle}{
\fancyhf{}
\lhead{\colorbox{scipostblue}{\bf \color{white} ~SciPost Physics Lecture Notes }}
\rhead{{\bf \color{scipostdeepblue} ~Submission }}

\fancyfoot[C]{\textbf{\thepage}}
}

\newcommand{\bra}[1]{\ensuremath{\langle#1\rvert}}
\newcommand{\ket}[1]{\ensuremath{\lvert#1\rangle}}

\newcommand{\dd}{\ensuremath{\text{d}}}

\newcommand{\mb}[1]{\ensuremath{\mathbf{#1}}} 

\begin{document}

\pagestyle{SPstyle}

\begin{center}{\Large \textbf{\color{scipostdeepblue}{
Analogue Hawking radiation in nonlinear quantum optics
}}}\end{center}

\begin{center}\textbf{
Isaac Bernal\textsuperscript{1$\star$}, 
Miguel A. Cortés-Ortiz\textsuperscript{2$\dagger$} and
David Bermudez\textsuperscript{2$\ddagger$}
}\end{center}

\begin{center}
{\bf 1} Escuela Superior de Física y Matemáticas, Instituto Politécnico Nacional, U. P. Adolfo López Mateos, Edificio 9, Zacatenco, 07730 Ciudad de México, Mexico
\\
{\bf 2} Department of Physics, Cinvestav, A.P. 14-740, 07000 Ciudad de México, Mexico
\\[\baselineskip]
$\star$ \href{mailto:ibernalr1900@alumno.ipn.mx}{\small ibernalr1900@alumno.ipn.mx}\,,\quad
$\dagger$ \href{mailto:miguel.cortes@cinvestav.mx}{\small miguel.cortes@cinvestav.mx}\,,\\
$\ddagger$ \href{mailto:david.bermudez@cinvestav.mx}{\small david.bermudez@cinvestav.mx}
\end{center}

\section*{\color{scipostdeepblue}{Abstract}}
\textbf{\boldmath{%
The Hawking effect can be understood as a broad kinematic phenomenon associated with mode behavior near a horizon. While astrophysical black holes produce one specific realization of this radiation, this perspective inspires extensive theoretical and experimental efforts to create event horizons in diverse physical systems  to observe the resulting analogue Hawking emission. One of the most successful realizations is the fiber-optical analogue, based on nonlinear quantum optics. In these notes, we introduce and motivate this system while outlining the theoretical concepts underlying the gravitational  analogy. Finally, we review key experiments and discuss their impact on the field.
}}

\vspace{\baselineskip}

\noindent\textcolor{white!90!black}{%
\fbox{\parbox{0.975\linewidth}{%
\textcolor{white!40!black}{\begin{tabular}{lr}%
  \begin{minipage}{0.6\textwidth}%
    {\small Copyright attribution to authors. \newline
    This work is a submission to SciPost Physics Lecture Notes. \newline
    License information to appear upon publication. \newline
    Publication information to appear upon publication.}
  \end{minipage} & \begin{minipage}{0.4\textwidth}
    {\small Received Date \newline Accepted Date \newline Published Date}%
  \end{minipage}
\end{tabular}}
}}
}


\vspace{10pt}
\noindent\rule{\textwidth}{1pt}
\tableofcontents
\noindent\rule{\textwidth}{1pt}
\vspace{10pt}


\section{Introduction}\label{sec.intro}
The two most successful theories of the 20th century were quantum theory \cite{Cohen.2019}---our most fundamental dynamical framework---and general relativity\cite{Misner.2017}---our most successful theory of gravity. However, efforts to unify them into a hypothetical quantum theory of gravity \cite{Ziaeepour.2022} remain unsuccessful. Despite this challenge, researchers utilize existing tools to probe the interface of these domains. In this case, a robust approach is quantum field theory on curved spacetimes (QFTCS) \cite{Birrell.1982,Wald.1994,Mukhanov.2007}, which treats the gravitational background as fixed and classical---not quantum--- while we analyze the behavior of test or probe quantum fields on top of it.

While QFTCS remains an approximation, it predicts several new and unexpected phenomena arising from the mixing of positive and negative frequency modes. Examples include the cosmological creation of particles (the Parker effect \cite{Parker.1968}), the acceleration radiation (the Unruh effect \cite{Unruh.1976}), and the black hole radiation (the Hawking effect\cite{Hawking.1974}). This work focuses on the latter.

In 1974, Stephen Hawking \cite{Hawking.1974} applied the QFTCS paradigm to black holes, modeling a classical black hole background interacting with a quantum field. Under these conditions, the black hole emits thermal radiation, now known as Hawking radiation (HR), with a temperature defined by
\begin{equation}
	k_\text{B}T=\frac{\hbar c^3}{8\pi GM},
\end{equation}
where $k_\text{B}$ is the Boltzmann constant, $\hbar$ is the reduced Planck's constant, $c$ is the speed of light, $G$ is the gravitational constant, and $M$ is the black hole mass. This phenomenon may provide critical insights into how to quantize gravity \cite{Reiner.2010}. Even more, multiple independent derivations reinforce the physical validity of the Hawking effect \cite{Belgiorno.2020}. However, all of these derivations rely on dubious assumptions \cite{Helfer.2003}, specifically the use of exponentially small lengths where the QFTCS paradigm likely breaks down. While observations should ideally resolve this contradiction, the effect remains too faint for detection in astrophysical settings using current or foreseeable technology \cite{Helfer.2004}.

\begin{figure}
    \centering
    \includegraphics[width=0.5\linewidth]{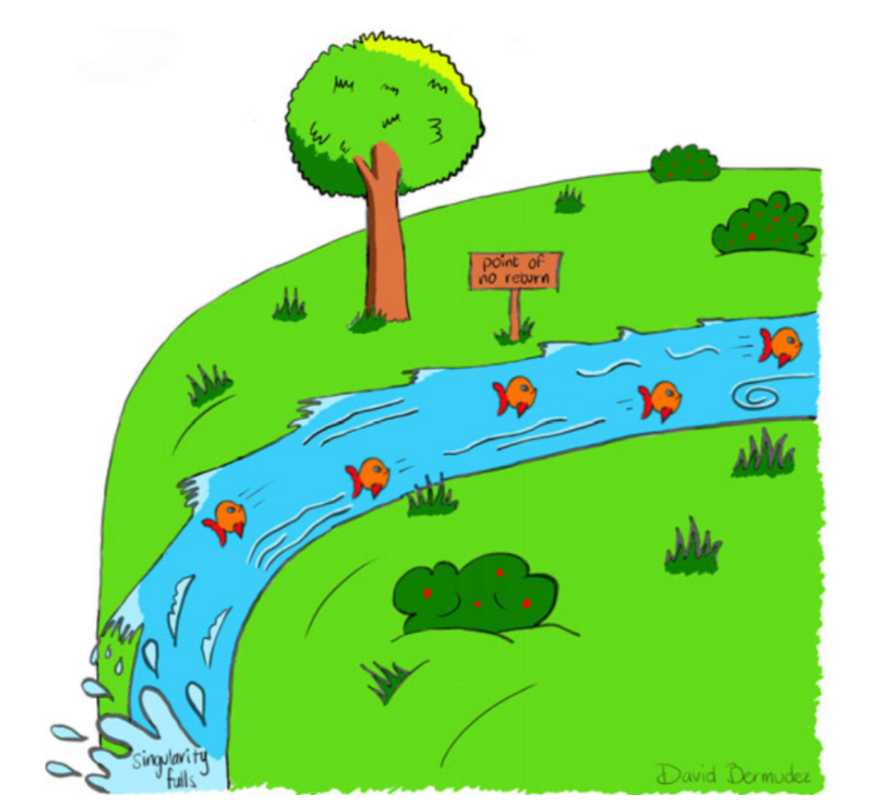}
    \caption{Schematic representation of the sonic analogue. The fishes play the role of sound waves. The river flow $v(x)$ changes as its slope changes. Image taken from \cite{Aguero.2020}.}
    \label{fig.Fish}
\end{figure}

These limitations initially suggested that the astrophysical Hawking effect would remain beyond experimental confirmation. However, in 1981 William Unruh \cite{Unruh.1981} demonstrated that the same effect occurs in a classical background of a moving fluid interacting with a quantum field of sound waves on top of it. This system is now known as a sonic black hole\footnote{Originally termed as a "dumb hole", dumb as in mute in high English. The name was later changed to avoid ambiguity.}, see Fig. \ref{fig.Fish}. This work reveals that the Hawking effect originates not from gravity itself, but from the kinematics of modes around a horizon\cite{Unruh.1995}. Recreating these kinematic conditions in other physical systems enables the generation of an analogue of Hawking radiation. This field---termed analogue gravity---replicates phenomena usually related to gravity in diverse analogue systems. We refer the reader to an extensive review \cite{Barcelo.2011} and its updated preprint \cite{Barcelo.2024} for further detail. 

Despite its significance, Unruh's work remained largely overlooked for a few years, a period known as the lost decade. In the 1990s, Ted Jacobson and others \cite{Jacobson.1991,Jacobson.1996} revitalized the theory, establishing the theoretical foundations of analogue gravity and the necessary conditions to reproduce the Hawking effect. By the 2000s, the scientific community initiated a search for experimental platforms capable of measuring this emission. Subsequent proposals and successful demonstrations spanned various systems, including water tanks \cite{Weinfurtner.2011, Rousseaux.2010, Rousseaux.2024}, Bose-Einstein condensates\cite{Garay.2000, Steinhauer.2010}, fluids of light \cite{Jacquet.2020_2}, and fiber optics \cite{Philbin.2008, Faccio.2012, Konig.2014, Drori.2019, Felipe-Elizarraras.2022}. This work focuses on the latter.

In 2008, the groups of Friedrich König and Ulf Leonhardt reported the first theoretical and experimental investigation into the fiber-optical analogue of Hawking radiation \cite{Philbin.2008}. In this context, the so-called resonant condition replaces the ubiquitous phase-matching conditions. The requirements to increase the efficiency of resonant effects---inducing analogue Hawking radiation---put them into a new regime of extreme nonlinear optics (XNLO). This requires ultra-short pulses (sub-100 fs) at intensities surpassing the fundamental soliton value. Since then, the study of the fiber-optical analogue of Hawking radiation has grown into an active research field with significant theoretical and experimental contributions \cite{Rubino.2012, Amiranashvili.2016, Baak.2025}, as we describe in these notes. 

The remainder of these notes is organized as follows. Section \ref{sec.preliminaries} provides the preliminaries and motivation, addressing the role of gravity, dispersion, and the so-called trans-Planckian problem. Section \ref{sec.movingmedia} analyzes light in moving media and demonstrates that the corresponding wave equation in moving media defines an effective geometry. Section \ref{sec.propagation} examines the pulse propagation equations used to predict and model the fiber-optical Hawking experiments. Section \ref{sec.timeline} provides a timeline of fiber-optical experiments from 2008 to the present. Finally, Section \ref{sec.conclusions} offers concluding remarks on the future of the field.

\section{Preliminaries}\label{sec.preliminaries}

\subsection{Black star}\label{sec.blackstar}
In 1783, geologist John Michell proposed the concept of a black star \cite{Michell.1783}. Within the framework of classical mechanics, this idea relies on two assumptions: light possesses a finite velocity and light is subject to gravity. Under these conditions, a sufficiently massive star exerts a gravitational attraction strong enough that its escape velocity matches the speed of light. Such a star would trap all emitted light, rendering it a black star.

\begin{figure}
	\centering
	\includegraphics[width=0.3\linewidth]{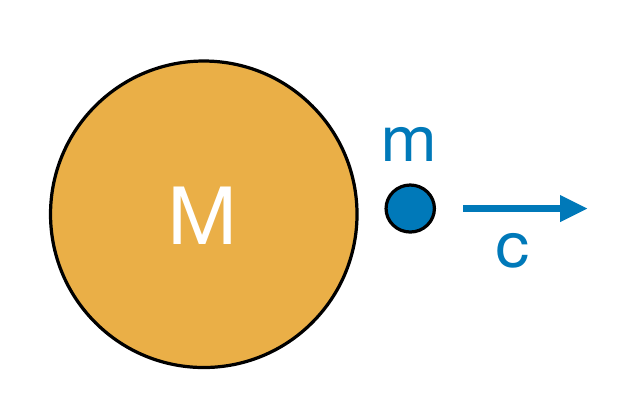}
	\caption{A diagram of a black star, where $M$ is the mass of the black star and $m$ and $c$ are the mass and velocity of the hypothetical light particle.}
	\label{fig.blackstar}
\end{figure}

Let us study the energy of a particle of mass $m$ moving away from the star. Its kinetic and potential energies are
\begin{equation}
	K =\frac{1}{2}mv^2,\qquad 
	U =-\frac{GMm}{r},
\end{equation}
where $G$ is the gravitational constant and $M$ is the stellar mass.

We want to consider the energy balance when we give the particle just the right amount of kinetic energy to overcome the gravitational potential energy of the star. Let $v_e$ be the escape velocity and  $r_e$ the escape radius. Setting the escape velocity equal to the speed of light $v_e=c$ and $r=r_e$, and the energy balance $K=-U$ becomes
\begin{equation}
	m\frac{c^2}{2}=\frac{GMm}{r_\mathrm{e}}.
\end{equation}
Solving for $r_\mathrm{e}$ yields
\begin{equation}
	r_\mathrm{e}=\frac{2GM}{c^2}, \label{GR}
\end{equation}
which recovers the gravitational (or Schwarzschild) radius obtained by solving the Einstein field equations for a spherically symmetric mass \cite{Schwarzschild.1916}.

We then calculate the acceleration at $r_\mathrm{e}$, denoted $a_\mathrm{e}$, from Newton's second law
\begin{equation}
	F=ma=\frac{GMm}{r^2}.
\end{equation}
Solving for $a_\mathrm{e}$ at $r=r_\mathrm{e}$, we find
\begin{equation}
	a_\mathrm{e}=\frac{GM}{r_\mathrm{e}^2}=\frac{GMc^4}{4G^2M^2}=\frac{c^4}{4GM}. \label{SG}
\end{equation}

This expression is also the same value as the surface gravity at the horizon in the Schwarzschild solution. The surface gravity at the horizon of a black hole is the acceleration needed to stay fixed at the horizon.

Finally, we assume the existence of the Unruh effect \cite{Unruh.1976}: the purely kinematic effect wherein an accelerating observer perceives a thermal spectrum of created particles. The Unruh temperature is defined as
\begin{equation}
	k_\text{B}T=\frac{\hbar a}{2\pi c},\label{Unruh}
\end{equation}
with $a$ being the acceleration. Substituting our calculated value of $a_\mathrm{e}$ for the black star into this expression yields
\begin{equation}
	k_\text{B}T=\frac{\hbar c^3}{8\pi G M}. \label{Hawking}
\end{equation}
This result matches the celebrated prediction by Stephen Hawking in his seminal paper in 1974 \cite{Hawking.1974}. In his original study, Hawking applied quantum field theory in curved spacetimes (QFTCS) to the collapsing geometry of a Schwarzschild black hole. Although our heuristic \textit{derivation} employs technically incorrect assumptions, the results still prompt a fundamental question: If the Hawking effect appears to be so robust as to almost emerge solely from Newtonian gravity, how could it possibly distinguish between competing theories of quantum gravity? For a discussion on this topic we refer the reader to Ref. \cite{Leonhardt.2010}.

\subsection{Black holes}\label{sec.bh}
Black holes are among the most captivating predictions of modern physics, perhaps for three rather different reasons.

To the imagination, BHs are profound. Their gravitational pull is so intense that nothing, not even light, can escape after crossing the boundary defined by the event horizon. Popular culture often depicts black holes as dangerous objects that may swallow everything in their way; however, black holes are not that dramatic. Most are spatially compact relative to astronomical distances, meaning they rarely interact with distant matter, rendering their presence in the universe largely unobtrusive.

To physicists, black holes are fascinating because they are, in a sense, simple. A non-rotating, uncharged black hole is described by the Schwarzschild solution to the Einstein field equations, with a spacetime geometry represented by the line element
\begin{equation}
\dd s^2=-\left(1-\frac{2GM}{c^2r}\right)c^2\dd t^2+\left(1-\frac{2GM}{c^2r}\right)^{-1}\dd r^2+r^2\dd\Omega^2.
\end{equation}
This metric exhibits two central features: A singularity at $r=0$, where the spacetime curvature diverges and general relativity breaks down, and an event horizon at the Schwarzschild radius
\begin{equation}
r_\text{S}=\frac{2GM}{c^2}.
\end{equation}
This radius acts as a boundary of no return, separating the exterior universe from the hidden interior. Remarkably, an observer freely falling through the horizon experiences no local anomalies; however, the horizon marks the global boundary beyond which signals can never reach null infinity.

Finally, black holes also serve as critical theoretical laboratories for unifying gravity and quantum theory. Near the horizon, we must describe quantum fields using the QFTCS framework, as the horizon reshapes the very notion of particle excitations. Field modes undergo significant distortion: outgoing modes that escape to infinity become entangled with partner modes falling toward the singularity. This mixing of modes across the horizon drives Hawking radiation and underpins the complexities of black hole thermodynamics and the information paradox \cite{Hawking.1976}.

\begin{figure}
	\centering
	\includegraphics[width=0.5\linewidth]{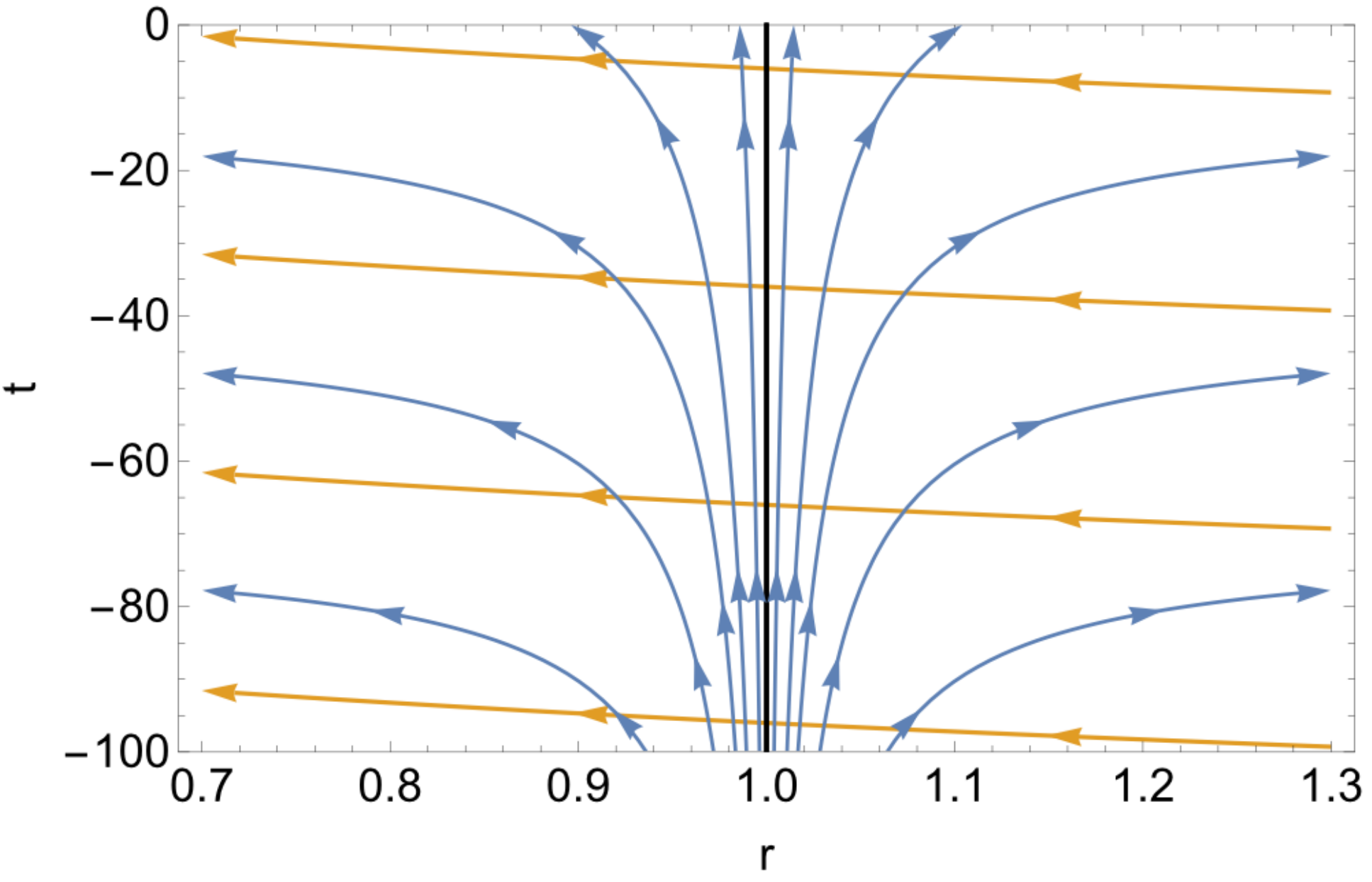}
	\caption{Ingoing and outgoing modes around a BH horizon $r=r_{\text{S}}$. Ingoing modes (yellow) see nothing special at $r_{\text{S}}$. Outgoing modes are trapped, for $r\leq r_{\text{S}}$, inside the event horizon.}
	\label{fig.lightcones}
\end{figure}

\subsection{Hawking radiation}\label{sec.hr}
Hawking radiation (HR) is the prediction that astrophysical black holes (BHs) emit thermal radiation due to quantum effects. The scientific community considers it a robust phenomenon, as multiple independent mathematical derivations consistently yield the same result: a quasi-stationary BH radiates a thermal spectrum with Hawking temperature \cite{Hawking.1974}
\begin{equation}
	k_\text{B}T_\text{H}=\frac{\hbar c^3}{8\pi GM}.
	\label{TH}
\end{equation}
The radiated power $P$ of a black body is given by the Stefan-Boltzmann law
\begin{equation}
	P=A\sigma T_{H}^4,
	\label{BBP}
\end{equation}
where $A$ is the surface area of the black hole and $\sigma$ is the Stefan--Boltzmann constant, defined as
\begin{equation}
	\sigma=\frac{\pi^2k_{\text{B}}^4}{60c^2\hbar^3}.
	\label{SB}
\end{equation}

This effect, however, is extremely faint and remains difficult to detect with our current or foreseeable technology. To illustrate the magnitude of the HR emission, we consider a BH in terms of the solar mass $\mathrm{M}_\odot\simeq1.989\times10^{30}\text{ kg}$. According to Eq. \eqref{TH}, the temperature is
\begin{equation}
	T=\frac{\hbar c^3}{8\pi G k_\text{B} M_\odot}\frac{M_\odot}{M}\simeq6.168\times 10^{-8}\text{ K}\cdot\frac{M_\odot}{M}.
\end{equation}
Similarly, using Eq. \eqref{BBP}, we find the power to be
\begin{equation}
	P=\frac{\hbar c^6}{15\cdot 2^{10}\pi G^2M_{0}^2}\frac{M_{0}^2}{M^2}\simeq9\times10^{-29}\text{ W}\cdot\frac{M_{0}^2}{M^2}.
\end{equation}

\subsection{Black hole of minimal mass}\label{sec.minimal}
Equation \eqref{TH} shows that the Hawking temperature depends inversely on the BH mass $M$; consequently, more massive BHs exhibit lower temperatures. Observations typically distinguish between two categories of BHs: (1) supermassive BHs, with masses reaching $\sim 10^{10}\mathrm{M}_\odot$, which possess temperature far too low to detect HR, and (2) stellar BHs, typically with masses of 3--20 $\mathrm{M}_\odot$ formed by gravitational collapse. The minimal mass required to collapse into a BH depends on the stellar equation of state. For instance, the Chandrasekhar limit establishes a minimum mass of approximately $1.44\,\mathrm{M}_\odot$\cite{Chandrasekhar.1931} for white dwarfs. For neutron stars, the Tolman--Oppenheimer--Volkoff limit suggests a minimum mass of 2--2.5 $\mathrm{M}_\odot$; such objects would be even colder than their white dwarf counterparts.

Substituting the Chandrasekhar limit into Eq. \eqref{GR} yields an estimated gravitational radius of
\begin{equation}
	r_\text{S}=\frac{2G(1.44\,\mathrm{M}_\odot)}{c^2}\simeq 4.3 \text{ km}.
\end{equation}
The corresponding Hawking temperature, calculated via Eq. \eqref{Hawking}, is
\begin{equation}
	T_\text{max} \simeq 4.3 \times 10^{-8} \text{ K}.
\end{equation}
Because the Hawking temperature is inversely proportional to the mass, the minimum collapse mass sets an upper bound on the temperature of stellar BHs.

The maximum BH temperature $T_\text{max}$ is several orders of magnitude smaller than the cold temperature in outer space dictated by the cosmic microwave background (CMB) at $\sim2.725$ K \cite{Kotani.2025}. Because more massive black holes are even colder, directly testing Hawking's prediction using astrophysical black holes presents an immense challenge, though it remains an active area of research \cite{Abedi.2023, Cacciapaglia.2024}.

Hawking himself immediately realized that experimental confirmation would require the existence of a third type of BH: primordial or microscopic BHs \cite{Hawking.1974}, with masses ranging from Planck's mass ($M_\text{P}\sim 10^{-5}$ g) to 1 $M_{\odot}$ \cite{Carr2020}. Another possibility, which we describe in Section \ref{sec.movingmedia}, is to recreate the kinematic characteristics of an event horizon using interactions other than gravity in systems that are more accessible to experiments \cite{Unruh.1981}.

\subsection{Trans-Planckian problem}\label{sec.TPP}
A rigorous derivation of the Hawking effect leads to waves with wavelengths $\lambda$ reaching future null infinity. However, tracing these modes backwards in time towards the horizon reveals an exponential blueshift. These waves quickly reach wavelengths that surpass the Planck length  $\ell_\text{P}\simeq 1.6\times 10^{-35}$ m, entering the regime where $\lambda<\ell_\text{P}$. In this region, the semiclassical approximation is no longer reliable \cite{Helfer.2003}, yet the standard derivation seems to depend precisely on this extrapolation. While the community often considers Hawking radiation one of the most robust predictions in QFTCS, it nonetheless rests on these dubious assumptions \cite{Helfer.2003}. 

In astrophysics, a yet unknown physical mechanism must regularize waves close to the horizon. Conversely, in analogue horizons recreated in different media, the regularization mechanism is known, it is dispersion, the dependence of the velocity of a wave on its frequency or wavenumber $c=c(\omega)$. Its inclusion limits redshifting or blueshifting at the BH or WH horizon, respectively.

In dispersive media, the concept of a horizon is no longer unique; the phase and group velocities
\begin{equation}
	v_{\textbf{p}}=\frac{\omega}{k},\qquad v_{\textbf{g}}=\frac{\dd\omega}{\dd k}, \label{vpvg} 
\end{equation}
do not coincide. One can then identify a phase-velocity horizon, defined by the condition $v_{\text{p}}=c$, which defines the relevant mode frequencies, and a group-velocity horizon $v_{\text{g}}=c$, which determines the kinematic blockage of wave packets. Even more, different frequencies may experience different effective horizons, leading to what is known as a fuzzy horizon, i.e., a different horizon location for each frequency \cite{Lang.2020}.

\subsection{Toy model of spacetime dispersion}\label{sec.toy}
\begin{figure}
	\centering
	\includegraphics[width=0.45\linewidth]{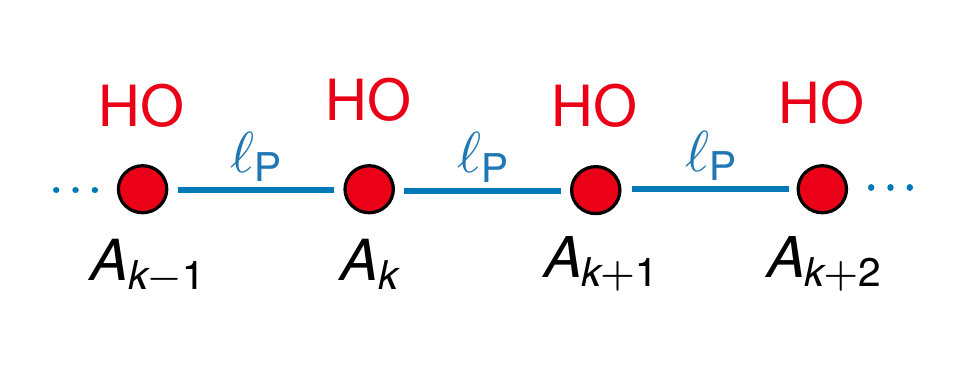}
	\caption{Toy model of spacetime dispersion with effective harmonic oscillators $\hat{A}_k$ at a distance $\ell_{\text{P}}$.}
	\label{Fig.Toy}
\end{figure}

If spacetime is discrete at a fundamental level, the first consequence to arise would be dispersion. To study this phenomenon, let us consider a toy model consisting of a 1-dimensional chain of harmonic oscillators (HOs) labeled by $k$, separated by a distance $\ell_\text{P}$, and described by an electromagnetic (EM) potential $\hat{A}_k\dag$ (see Fig. \ref{Fig.Toy}).

Expanding the EM potential to first neighbors
\begin{equation}
	\frac{\partial^2\hat{A}_k}{\partial t^2}=\kappa(\hat{A}_{k+1}-\hat{A}_k)+\kappa(\hat{A}_{k-1}-\hat{A}_k)= \hat{A}(x_{k+1})-2\hat{A}(x_k)+\hat{A}(x_{k-1}),
\end{equation}
considering the continuous limit $\hat{A}_k(t)\rightarrow\hat{A}(x_k,t)$. Taking Taylor's expansion of $\hat{A}(x,t)$ around $x_0$ up to 4th order, we have
\begin{align}
	\hat{A}(x,t)=&\hat{A}(x_0,t)+\hat{A}^{(1)}(x_0,t)(x-x_0)+\frac{\hat{A}^{(2)}}{2}(x_0,t)(x-x_0)^2\nonumber\\
	&+\frac{\hat{A}^{(3)}}{6}(x_0,t)(x-x_0)^3+ \frac{\hat{A}^{(4)}}{24}(x_0,t)(x-x_0)^4,
\end{align}
where $\hat{A}^{(m)}$ is the $m$th-order derivative with respect to $x$.

For example, consider the case $x_0=x_k$, the above equation becomes
\begin{align}
	\hat{A}(x,t)=&\hat{A}(x_k,t)+\hat{A}^{(1)}(x_k,t)(x-x_k)+\frac{\hat{A}^{(2)}}{2}(x_k,t)(x-x_k)^2\nonumber\\
	&+\frac{\hat{A}^{(3)}}{6}(x_k,t)(x-x_k)^3+\frac{\hat{A}^{(4)}}{24}(x_k,t)(x-x_k)^4.
\end{align}

We need to calculate $\hat{A}(x_{k+1})-2\hat{A}(x_k)+\hat{A}(x_{k-1})$. Separating each power as
\begin{alignat}{2}
	&\text{0th-order}: &\quad  \hat{A}(0)-2\hat{A}(0)+\hat{A}(0)&=0,\\
	&\text{1st-order}: &\quad \hat{A}'(0)(\ell_\text{P}+2\cdot0-\ell_\text{P})&=0,\\
	&\text{2nd-order}: &\quad \frac{\hat{A}''(0)}{2}(\ell_\text{P}^2+0+\ell_\text{P}^2)&=\ell_\text{P}^2\hat{A}''(\ell_\text{P}),\\
	&\text{3rd-order}: &\quad \frac{\hat{A}'''(0)}{6}(\ell_\text{P}^3+0-\ell_\text{P}^3)&=0,\\
	&\text{4th-order}: &\quad \frac{\hat{A}^{(4)}(0)}{24}(\ell_\text{P}^4+0+\ell_\text{P}^4)&=\frac{\ell_\text{P}^4}{12}A^{(4)}(\ell_\text{P}).
\end{alignat}
The resulting equation is
\begin{equation}
	\frac{\partial^2\hat{A}}{\partial t^2}=c^2\left(\frac{\partial^2 \hat{A}}{\partial x^2}+\frac{\ell_\text{P}^2}{12}\dfrac{\partial^4\hat{A}}{\partial x^4}\right), \label{WEEMFET}
\end{equation}
where we define $c^2=k\ell_\text{P}^2$.

Equation \eqref{WEEMFET} resembles the wave equation for the electromagnetic field
\begin{equation}
	\frac{\partial H}{\partial x}+\frac{\partial D}{\partial t}=\frac{1}{\mu_0}\frac{\partial B}{\partial x}+\epsilon_0\frac{\partial E}{\partial t}=\frac{1}{\mu_0}\frac{\partial^2A}{\partial t^2}-\epsilon_0\frac{\partial^2A}{dt^2}=0, \label{WEA}
\end{equation}
plus an additional term that depends on $\hat{A}^{(4)}$.

To find the dispersion relation, we take a plane wave Ansatz
\begin{equation}
	\hat{A}=\hat{\mathcal{A}} e^{ikx-i\omega t},
\end{equation}
then
\begin{align}
	(i\omega)^2\hat{A}&=c^2\left[ (ik)^2+\frac{\ell_\text{P}^2}{12}(ik)^4\right]\hat{A}\\
	\omega(k)&=ck\sqrt{1-\frac{k^2}{k_0^2}},
\end{align}
with $k_0=\sqrt{12}/\ell_\text{P}$. This last equation proves that spacetime discretization appears as dispersion. In fact, dispersion modifies both the phase and group velocities.

For the phase velocity
\begin{equation}
	v_p(k)=\frac{\omega(k)}{k}=c\sqrt{1-\frac{k^2}{k_0^2}},
\end{equation}
and for the group velocity
\begin{align}
	v_g(k)&=\frac{\partial\omega(k)}{\partial k}=\frac{\partial}{\partial k}\left(ck\sqrt{1-\frac{k^2}{k_0^2}}\right)=c\sqrt{1-\frac{k^2}{k_0^2}}\left( 1-\frac{k^2}{k_0^2-k^2}\right).
\end{align}

\subsection{The analogue gravity paradigm}\label{sec.paradigm}
The main idea of analogue gravity is that the kinematics of classical or quantum excitations on a curved spacetime geometry can be reproduced in different analogue systems with interactions different from gravity. In fact, analogue spacetime has recently been suggested as a more suitable name \cite{Barcelo.2024}. To get the same paradigm as QFTCS, we separate a classical background field that determines an effective curved spacetime and a quantum excitation or probe field propagating on top of this effective curved spacetime that mimic the kinematics of fields in general relativity \cite{Philbin.2008}.

\begin{figure}
	\centering
	\includegraphics[width=0.6\linewidth]{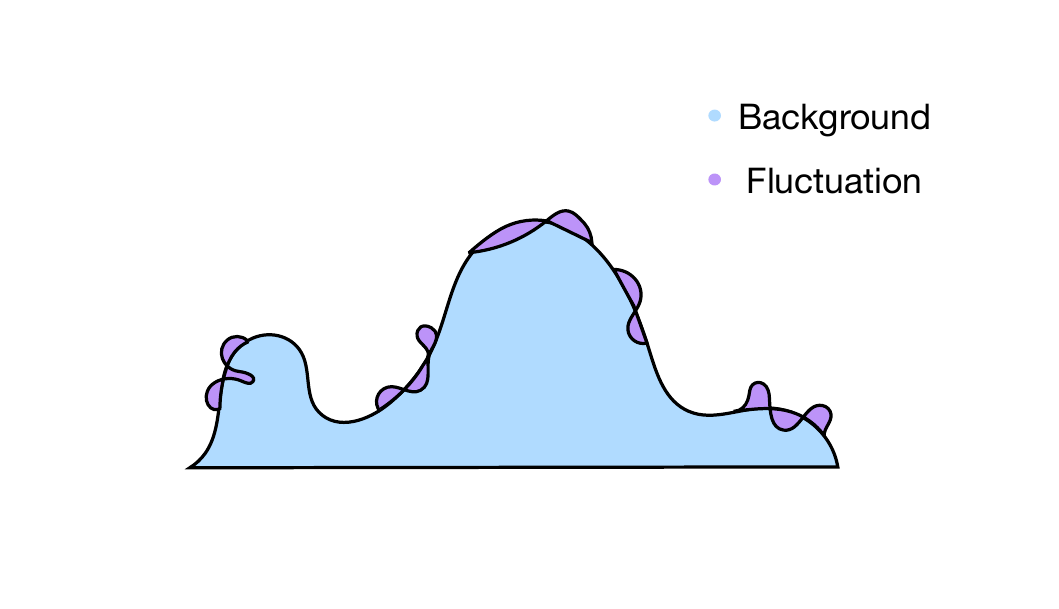}
	\caption{Diagram of the system showing the classical or quantum fluctuations over a classical background.}
	\label{fig:DAG}
\end{figure}

Some of the effects associated with QFTCS in which analogue gravity is making significant progress in the laboratory include Hawking radiation (HR), the dynamical Casimir effect \cite{Vezzoli.2019}, superradiance \cite{Patrick.2017, Braidotti.2020}, cosmological creation of particles \cite{Fedichev.2004, Tolosa-Simeon.2024}, the Hubble tension \cite{Berechya.2021} and false vacuum decay \cite{Jenkins.2024}. 

The following notes will focus on HR, both as a proof of principle and because this effect has been the most widely studied in the context of analogue gravity. This includes several different physical systems, ranging from water waves \cite{Weinfurtner.2011} to Bose-Einstein condensates (BECs) \cite{Garay.2000} and nonlinear optical systems \cite{Bermudez.2016pra}.

\section{Light in moving media}\label{sec.movingmedia}
\subsection{Wave equation in moving media}
Let us turn to the study of how light transforms from one inertial frame to another, as this would be an essential tool to obtain the effective optical metric and develop analogue gravity in optics. We know that the causal velocity is $c$ in any inertial frame, and corresponds to the velocity of light in vacuum. Let us consider a fixed polarization in the $yz$-plane, propagation along $x$, and let $A$ be the vector potential component. 

Under these conditions---not in general---the value of $A$ in the lab frame $(x,t)$ is the same as $A'$ in the moving frame $(x',t')$. We can express the electric and magnetic field operators in terms of the vector potential component
\begin{alignat}{3}
\mathbf{E} &= -\frac{\partial \mathbf{A}}{\partial t},
 \qquad E &= -\frac{\partial A}{\partial t},
 \qquad E' &= -\frac{\partial A'}{\partial t'} \\
\mathbf{B} &= \nabla \times \mathbf{A},
 \qquad B &= \frac{\partial A}{\partial x},
 \qquad B' &= \frac{\partial A'}{\partial x'}
\label{BET}
\end{alignat}
Using the Lorentz transformations with velocity $u$ between the two frames
\begin{equation}
	x=\gamma(x'+ut'),\qquad t=\gamma \left( t'+\frac{u}{c^2}x' \right),
\end{equation}
where $\gamma$ is the Lorentz factor. We use the chain rule to express both $E'$ and $B'$ in terms of laboratory fields. For $E'$, we have
\begin{equation}
	\frac{\partial}{\partial t'}=\frac{\partial t}{\partial t'}\frac{\partial}{\partial t}+\frac{\partial x}{\partial t'}\frac{\partial}{\partial x}=\gamma\frac{\partial}{\partial t}+\gamma u\frac{\partial}{\partial x}=\gamma\left(\frac{\partial}{\partial t}+u\frac{\partial}{\partial x}\right). \label{deltatp}
\end{equation}
Then, we can express $E'$ as
\begin{equation}
	E'=-\frac{\partial A}{\partial t'}=-\gamma\left(\frac{\partial A}{\partial t}+u\frac{\partial A}{\partial x}\right)=\gamma(E-uB).
\end{equation}

Using an equivalent analysis for $\partial_{x'}$, 
\begin{align}
	\frac{\partial}{\partial x'}=\frac{\partial t}{\partial x'}\frac{\partial}{\partial t}+\frac{\partial x}{\partial x'}\frac{\partial}{\partial x}	=\gamma\frac{u}{c^2}\frac{\partial}{\partial t}+\gamma\frac{\partial}{\partial x}=\gamma\left(\frac{\partial}{\partial x}+\frac{u}{c^2}\frac{\partial}{\partial t}\right).\label{deltaxp}
\end{align}
Then, we can express $B'$ as
\begin{equation}
	B'=\frac{\partial A}{\partial x'}=\gamma\left( \frac{\partial A}{\partial x}+\frac{u}{c^2}\frac{\partial A}{\partial t}\right)=\gamma\left( B-\frac{u}{c^2}E \right).
\end{equation}

According to the second postulate of special relativity, the laws of physics are the same in any inertial frame, including those of electromagnetism. So taking the wave equation
\begin{equation}
	\nabla\times \mathbf{H}-\frac{\partial \mathbf{D}}{\partial t}=0,
\end{equation}
 in (1+1)D and transforming it to another reference frame
\begin{equation}
	\frac{\partial H}{\partial x}+\frac{\partial D}{\partial t}=0\quad\Rightarrow\quad\frac{\partial H'}{\partial x'}+\frac{\partial D'}{\partial t'}=0. \label{WE}
\end{equation}

Using eqs. \eqref{deltatp} and \eqref{deltaxp} to transform the primed derivatives in Eq. \eqref{WE} we obtain
\begin{equation}
	\frac{\partial H'}{\partial x'}+\frac{\partial D'}{\partial t'}=\gamma\frac{\partial}{\partial x}(H'+uD')+\gamma\frac{\partial}{\partial t}(D'+\frac{u}{c^2}H')=0.
\end{equation}

For this reason, we conclude that the fields $H$ and $D$ should transform as	
\begin{equation}
	H=\gamma(H'+uD'); \qquad D=\gamma\left( D'+\frac{u}{c^2}H' \right).\label{HDB}
\end{equation}

Now, we use the constitutive equations, which are valid in the rest frame of the medium, i.e. the laboratory frame,
\begin{equation}
	\mathbf{D}=\epsilon_0\epsilon\mathbf{E};\qquad\mathbf{B}=\mu_0\mu\mathbf{H},\label{ConE}
\end{equation}
Remembering that in vacuum $\epsilon=\mu=1$, we can rewrite the wave equation in terms of the vector potential $A$ as
\begin{equation}
	\frac{\partial H}{\partial x}+\frac{\partial D}{\partial t}=\frac{1}{\mu_0}\frac{\partial B}{\partial x}+\epsilon_0\frac{\partial E}{\partial t}=\frac{1}{\mu_0}\frac{\partial^2A}{\partial t^2}-\epsilon_0\frac{\partial^2A}{dt^2}=0, \label{WEA2	}
\end{equation}
the wave equation for the transformed quantities $A'$, $x'$, and $t'$ take the same form as the equation above.

We now turn to the derivation of the effective metric. To do this, we need to calculate the fields $\mathbf{D}$ and $\mathbf{H}$ in a comoving frame with velocity $u$. Inverting the expressions of $D$ to obtain $D'$ and $H'$
\begin{equation}
	D'=\gamma\left(D-\frac{u}{c^2}H\right), \qquad H'=\gamma(H-uD),\label{HP}
\end{equation}
Substituting these fields into the constitutive equations, we get
\begin{equation}
	D'+\frac{u}{c^2}H'=\epsilon_0\epsilon(E'+uB'),\qquad H'+uD'=\frac{1}{\mu_0\mu}\left(B'+\frac{u}{c^2}E'\right).
\end{equation}
Solving for $D'$ and $H'$ we get
\begin{equation}
	D'=\gamma^2\epsilon_0\left[\epsilon(E'+uB')-\frac{u}{\mu}\left(B'+\frac{u}{c^2}E'\right)\right], \quad H'=\gamma^2\epsilon_0\left[\frac{c^2}{\mu}\left(B'+\frac{u}{c^2}E'\right)-u\epsilon(E'+uB')\right],\label{CELFH}
\end{equation}
where we used the fact that $c^2=(\mu_0\epsilon_0)^{-1}$. 

Equations \eqref{CELFH} are the constitutive equations in the comoving frame. Since we already have expressions for $B'$ and $E'$ in terms of $A$ in Eq. \eqref{BET}, we can rewrite these last equations as
\begin{equation}
	D'=\gamma^2\epsilon_0\left[\left(\frac{u^2}{\mu c^2}-\epsilon\right)c\partial'_{0}+\left(\epsilon-\frac{1}{\mu}\right)u\partial'_1\right]A, \quad
	H'=\gamma^2\epsilon_0\left[\left(\epsilon-\frac{1}{\mu}\right)uc\partial'_{0}+\left(\frac{c^2}{\mu}-\epsilon u^2\right)\partial'_1\right]A,\label{CELHDA}
\end{equation}
where $\partial'_t=c\partial'_0$ and $\partial'_x=\partial'_1$.

\subsection{Field quantization}\label{sec.quantization}
Let us briefly review the quantization of the electromagnetic field, which is essential for a proper understanding of the optical analogue. To quantize the electromagnetic field, let us consider a linear polarization transverse to the propagation direction, such that the vector potential $\hat{\mathbf{A}}$ reduces to a scalar field $\hat{A}$. Then it can be expressed as a superposition of modes
\begin{equation}
	\hat{A} = \sum_k \left( A_k(\mathbf{r},t)\hat{a}_k + A_k^{*}(\mathbf{r},t)\hat{a}_k^{\dagger} \right),\label{AME}
\end{equation}
where $A_k$ denotes the mode functions, $k$ the mode index, and $\hat{a}$ and $\hat{a}^{\dagger}$ the mode operators. The operator $\hat{A}$ fully describes the quantum field of light. For a more in-depth review of the quantization of the EM field see Refs. \cite{Leonhardt.2010,Gerry.2023,Drummond.2014}.

Let us now define the quantum operators $\hat{D}$ and $\hat{H}$ in terms of Eq. \eqref{AME}, such that $H=\bra{\psi}\hat{H}\ket{\psi}$ and $D=\bra{\psi}\hat{D}\ket{\psi}$. Then we can write Eqs. \eqref{CELHDA} as
\begin{equation}
	\hat{D}=-\epsilon_0cg^{0\beta}\partial_\beta\hat{A},\qquad\hat{H}=-\epsilon_0c^2g^{1\beta}\partial_\beta\hat{A}, \label{QDH}
\end{equation}
using the matrix
\begin{equation}
	\displaystyle{g}^{\alpha\beta}=\gamma^2\begin{pmatrix}
		\displaystyle \epsilon-\frac{1}{\mu}\frac{u^2}{c^2}&\displaystyle \frac{u}{c}\left(\epsilon-\frac{1}{\mu}\right)\\
		\displaystyle \frac{u}{c}\left(\epsilon-\frac{1}{\mu}\right) &\displaystyle \epsilon\frac{u^2}{c^2}-\frac{1}{\mu}
	\end{pmatrix}.\label{IEM}
\end{equation}

Maxwell's equations are valid in both frames and they also govern the quantum world. This is
\begin{equation}
	\frac{\partial\hat{H}}{\partial x}+\frac{\partial\hat{D}}{\partial t}=0; \qquad \frac{\partial\hat{H'}}{\partial x'}+\frac{\partial\hat{D'}}{\partial t'}=0.
\end{equation}
Finally, using Eqs. \eqref{QDH} it follows that
\begin{equation}
	\frac{\partial\hat{H}}{\partial x}+\frac{\partial\hat{D}}{\partial t}=c\partial_0\hat{D}+\partial_1\hat{H}=-\epsilon_0c^2\partial_0g^{o\beta}\partial_\beta\hat{A}-\epsilon_0c^2\partial_1g^{1\beta}\partial_\beta\hat{A}=\partial_\alpha g^{\alpha\beta}\partial_\beta\hat{A}=0.\label{WEMM}
\end{equation}
This last equation is the wave equation in moving media.

In the case under study the laboratory spacetime is flat. We have shown that it is possible to rewrite the wave equation in terms of an inverse effective metric $g^{\alpha\beta}$ as in Eq. \eqref{IEM}. The field $\hat{A}$ propagates in an effective curved spacetime given by $g_{\alpha\beta}=(g^{\alpha\beta})$, such that

\begin{equation}
	g_{\alpha\beta}=\gamma^{2}\begin{pmatrix}
		\displaystyle \frac{1}{\epsilon}-\mu\frac{u^2}{c^2}&\displaystyle \left(\mu-\frac{1}{\epsilon}\right)\frac{u}{c}\\
		\displaystyle \left(\mu-\frac{1}{\epsilon}\right)\frac{u}{c}&\displaystyle \frac{u^2}{c^2}\frac{1}{\epsilon}-\mu
	\end{pmatrix}.
\end{equation}

We can impose the impedance-matching condition $\mu=\epsilon=u$ and define $n^2=\mu\epsilon$ to simplify the metric as

\begin{equation}
	g_{\alpha\beta}=\gamma^{2}\begin{pmatrix}
		\displaystyle 1-n^2\frac{u^2}{c^2}&\displaystyle (n^2-1)\frac{u}{c}\\
		\displaystyle (n^2-1)\frac{u}{c}&\displaystyle \frac{u^2}{c^2}-n^2
	\end{pmatrix}.\label{EM}
\end{equation}
This equation confirms that a moving medium induces an effective curved spacetime geometry.

\subsection{Hawking radiation in the optical analogue}\label{sec.opticalhr}

According to Unruh, a valid analogue requires a medium moving at the speed of the waves within that medium. In our case of interest, this is the speed of light inside an optical fiber---a dielectric--- that is, approximately 2c/3. This raises the question of whether such condition could even be satisfied in the laboratory. In fact, this condition can be realized and is commonly exploited in optical communication technologies \cite{Agrawal.2008cap7}.

\begin{figure}
	\centering
	\includegraphics[width=0.6\linewidth]{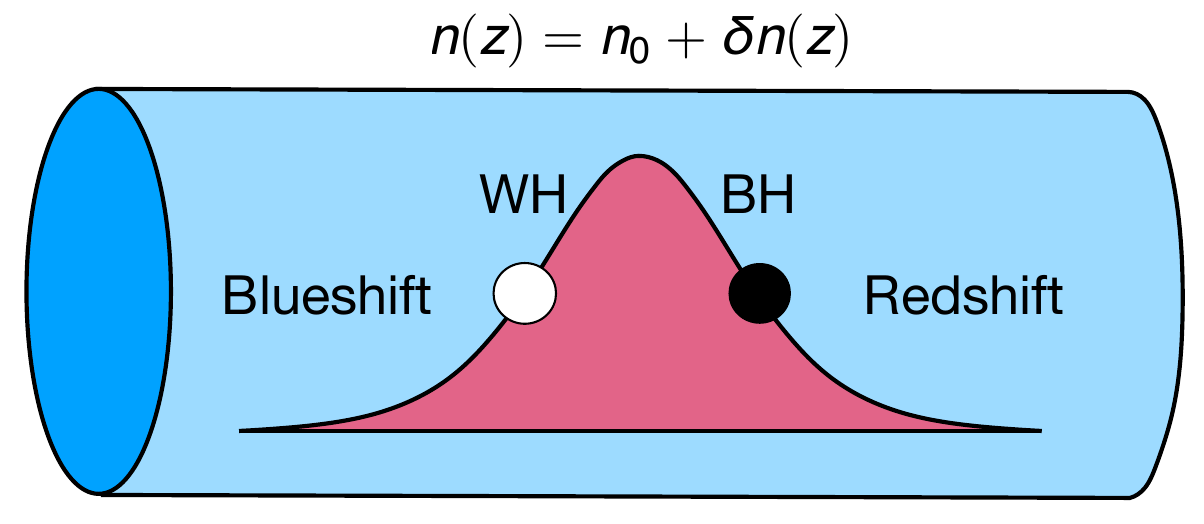}
	\caption{A refractive-index perturbation induced by a Kerr pulse propagating along the optical fiber. The leading and trailing edges of the perturbation act as analogue horizons: a WH horizon associated with a blueshift of probe modes, and a BH horizon associated with a redshift.}
	\label{fig.Kerr}
\end{figure}

The basic idea of this analogue is that a pulse in a medium with $\chi^{(3)}$ nonlinearity changes the refractive index due to the optical Kerr effect, and it establishes an effective curved spacetime (see Fig. \ref{fig.Kerr}). It assumes that light is only moving in the longitudinal axis of the fiber, therefore reducing the horizon physics to (1+1)D. 

\begin{figure}
	\centering
	\includegraphics[width=0.6\linewidth]{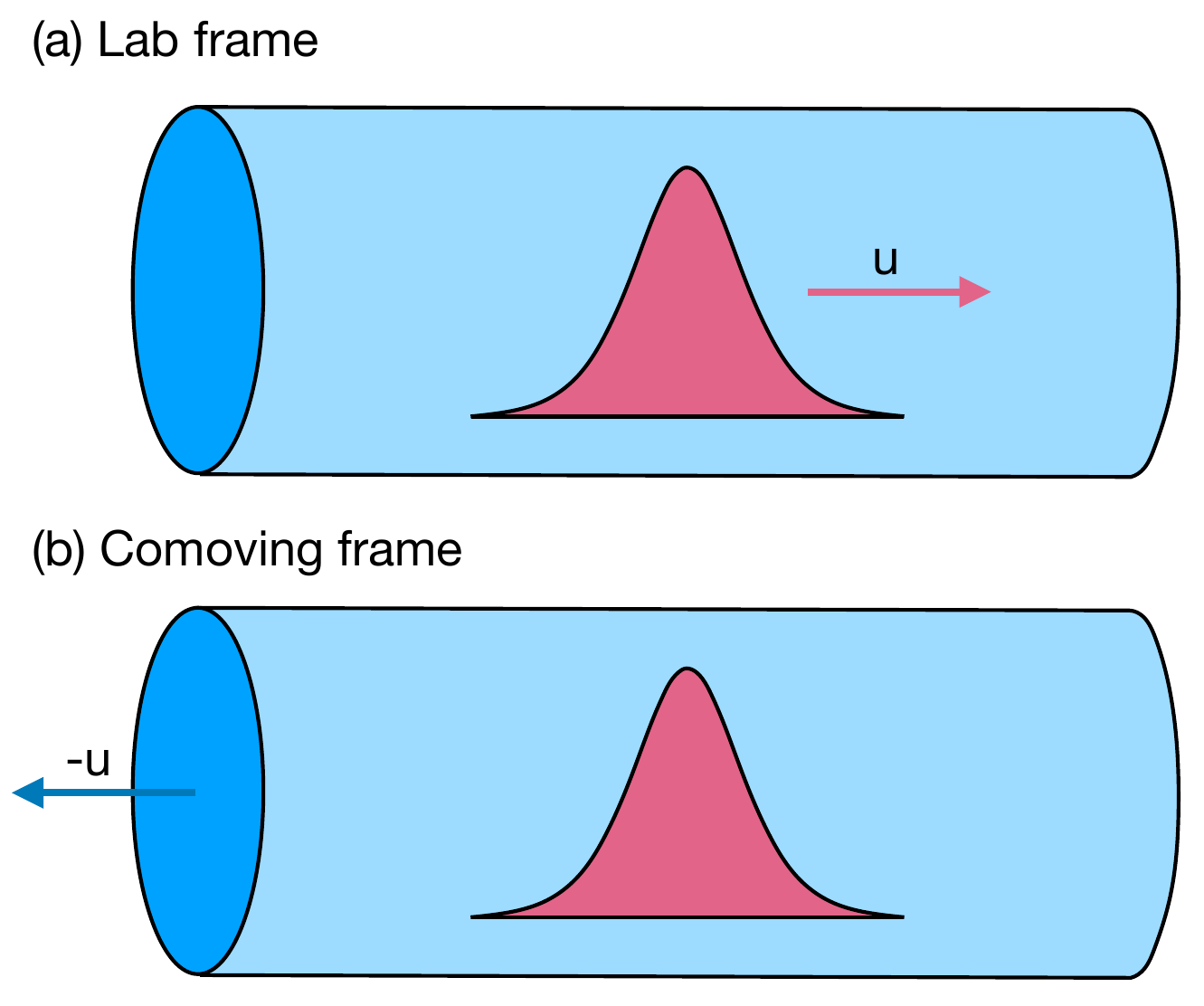}
	\caption{(a) An optical pulse traveling through a fiber seen from the laboratory frame. (b) A fiber traveling in the comoving frame of the optical pulse.}
	\label{fig.Frames}
\end{figure}

We consider a pulse moving with a group velocity $v_g=+u$ through an optical fiber with a refractive index $n_0(\omega)$. In the comoving frame, the optical fiber is moving with a velocity $-u$ (see Fig. \ref{fig.Frames}).
Considering the effective metric in Eq. \eqref{EM}, we impose the horizon condition $g_{00}=0$, then 
\begin{equation}
	1-n^2\frac{u^2}{c^2}=0\rightarrow u=\pm\frac{c}{n}=v_p,\label{vh}
\end{equation}
where $v_p$ is the phase velocity, and this condition determines the phase-velocity horizon.

Expanding to first order in the velocity profile at the horizon $x=x_h$
\begin{align}
	v_+=\alpha x, \qquad\alpha=\left. \frac{\dd v_+}{\dd x}\right|_{x=x_h}.
\end{align}
From the relativistic velocity addition formula,
\begin{equation}
	v_\pm=\frac{u\pm\frac{c}{n}}{1+\frac{u}{cn}}=\frac{\frac{un\pm c}{u}}{\frac{cn+u}{cn}}=\frac{unc\pm c^2}{cn+u}.
\end{equation}
In general, $u=u(x)$, $n=n(x)$, and $c=\text{const.}$, then
\begin{equation}
	\dfrac{\dd v_+}{\dd x}=\frac{(cn+u)c(u'n+un')-c(un+c)(cn'+u')}{(cn+u)^2}=\frac{c[c(n^2-1)u'+(u^2-c^2)n']}{(cn+u)^2}.\label{vd}
\end{equation}

Substituting the horizon condition in Eq. \eqref{vh} $u=-c/n$ into the derivative, we obtain for $\alpha$:
\begin{equation}
	\alpha=\left.\dfrac{\dd v_+}{\dd x}\right|_{x=x_h}=\frac{c^2(n^2-1)u'+c^2\left(\displaystyle\frac{1}{n^2}-1\right)n']}{\left(cn+\displaystyle\frac{c}{n}\right)^2}=\frac{(n^2-1)u'-\displaystyle\frac{1}{n^2}(n^2-1)}{\displaystyle\frac{1}{n^2}(n^2-1)^2}.
\end{equation}
Finally,
\begin{equation}
	\alpha=\frac{1}{1-n^{{-2}}}\left(u'-\frac{1}{n^2}n'\right)=\frac{1}{1-n^{-2}}\left( \frac{\dd u}{\dd x}-\frac{1}{n^2}\left. \frac{\dd n}{\dd x}\right)\right|_{x=x_h}.\label{alph}
\end{equation}

The case when $\alpha>0$ gives a black hole (BH), and when $\alpha<0$ gives a white hole (WH), the time reversal of a black hole. The WH solution is known for general relativity, but there is no known physical mechanism to form it in astrophysics. We focus on the BH case. 

The velocity can be linearized as $v(x) \simeq \alpha x$ close to the horizon, which allows us to define the coordinate
\begin{equation}
	t_+ = t - \int \frac{dx}{v_+}= t - \int \frac{dx}{\alpha x}= t - \frac{1}{\alpha}\ln |x| .
\end{equation}
This logarithmic relation implies that an exponential decrease in $x \to 0$ corresponds to a linear advance in $t_+$.
As a consequence, right-moving modes are increasingly compressed near the horizon.

The near-horizon dynamics leads to a mixing of left- and right-moving modes, which can be described by a Bogoliubov transformation as in the Unruh effect
\begin{align}
	A_1 &= A_R \cosh \zeta + A_L^\ast \sinh \zeta , \\
	A_2 &= A_L \cosh \zeta + A_R^\ast \sinh \zeta .
\end{align}
The relative amplitudes satisfy
\begin{equation}
	\frac{\beta}{\alpha} = \tanh \zeta = e^{-\nu \pi},
	\qquad
	\nu = \frac{\omega}{\alpha}.
\end{equation}
Together with the normalization condition
\begin{equation}
	|\alpha|^2 - |\beta|^2 = 1 ,
\end{equation}
which implies
\begin{equation}
	 \langle n \rangle=|\beta|^2=\frac{1}{e^{2\pi\omega/\alpha}-1}.
\end{equation}
Finally, by comparing with a thermal spectrum, we obtain the effective temperature
\begin{equation}
	\langle n \rangle =\frac{1}{e^{\hbar\omega/k_{B}T}-1},
\end{equation}
we obtain,
\begin{equation}
	k_{B}T=\frac{\hbar\alpha}{2\pi}.\label{OAHR} 
\end{equation}
This equation gives the effective temperature of the optical analogue of Hawking radiation. Recall that $\alpha$ is given in Eq. \eqref{alph} and depends on the rate of change of $u(x)$ and $n(x)$.

A BH horizon creates a two-mode squeezed vacuum that can be well approximated as an Einstein--Podolski--Rosen (EPR) pair
\begin{equation}
	\ket{\psi}=e^{ei\zeta\hat{K}_y}\ket{0,0},\label{EPR}
\end{equation}
where
\begin{equation} 
\hat{K}_y=\frac{i}{2}(\hat{a}_1\hat{a}_{2}-\hat{a}_{1}\dag \hat{a}_{2}\dag).
\end{equation}
We can see from this equation that the pair is entangled.

Then the reduced state on any side is a thermal state with the form
\begin{equation}
	\hat{\rho}=(1-e^{-\beta})\sum_{n=0}^{\infty}e^{n\beta}\ket{n}\bra{n}.
\end{equation}
Equation \eqref{alph} can be compared with similar equations in other analogue systems, such as water tanks and BECs. Next, we particularize to the fiber-optical case.

\subsection{HR in the fiber-optical case}\label{sec.fiberoptics}
In fiber optics, $u$ arises from the transformation to the comoving frame and is constant, $u\neq u(x)$; only the refractive index varies spatially $n=n(x)$. Then Eq. \eqref{alph} reduces to
\begin{equation}
	\alpha=\frac{1}{1-n^{-2}}\left( \frac{\dd u}{\dd x}-\frac{1}{n^2}\left. \frac{\dd n}{\dd x}\right)\right|_{x=x_h}
	=-\gamma^2\frac{c}{n^{2}}\left.\frac{\dd n}{\dd x}\right|_{x=x_h}
	=-\frac{c\gamma^2}{n^2}\left.\frac{\dd n}{\dd x}\right|_{x=x_h}
\end{equation}
with
\begin{equation}
	\gamma=\frac{1}{\sqrt{1-\frac{u^2}{c^2}}}=\frac{1}{\sqrt{1-n^{-2}}}.
\end{equation}
The calculation is performed in a comoving (Lorentz-boosted) frame $(x',t')$
\begin{equation}
\alpha\to\alpha', \qquad x\to x',\qquad t\to t'.
\end{equation}
In this frame, the effective acceleration is
\begin{equation}
	\alpha' = - \frac{c^2}{n^2}\,\frac{\dd n}{\dd x'} .\label{SGO}
\end{equation}
The corresponding derivatives transform according to
\begin{equation}
	\frac{\partial}{\partial t'} = \gamma\!\left(\frac{\partial}{\partial t}-u\frac{\partial}{\partial x}\right), \qquad \frac{\partial}{\partial x'} = \gamma\left(\frac{\partial}{\partial x}-\frac{u}{c^2}\frac{\partial}{\partial t}\right).
\end{equation}

In fiber-optics, it was discovered that it is more convenient to work in the retarded frame
$(\tau,\zeta)$, defined by
\begin{equation}
	\tau = t - \frac{x}{u}, \qquad \zeta = \frac{x}{u} , \label{coco}
\end{equation}
with inverse relations
\begin{equation}
	t = \tau+\frac{\zeta}{u}, \qquad x = u\zeta ,
\end{equation}
see Fig. \ref{fig.Frames} for a comparison between the laboratory and retarded frames.
\begin{figure}
	\centering
	\includegraphics[width=0.6\linewidth]{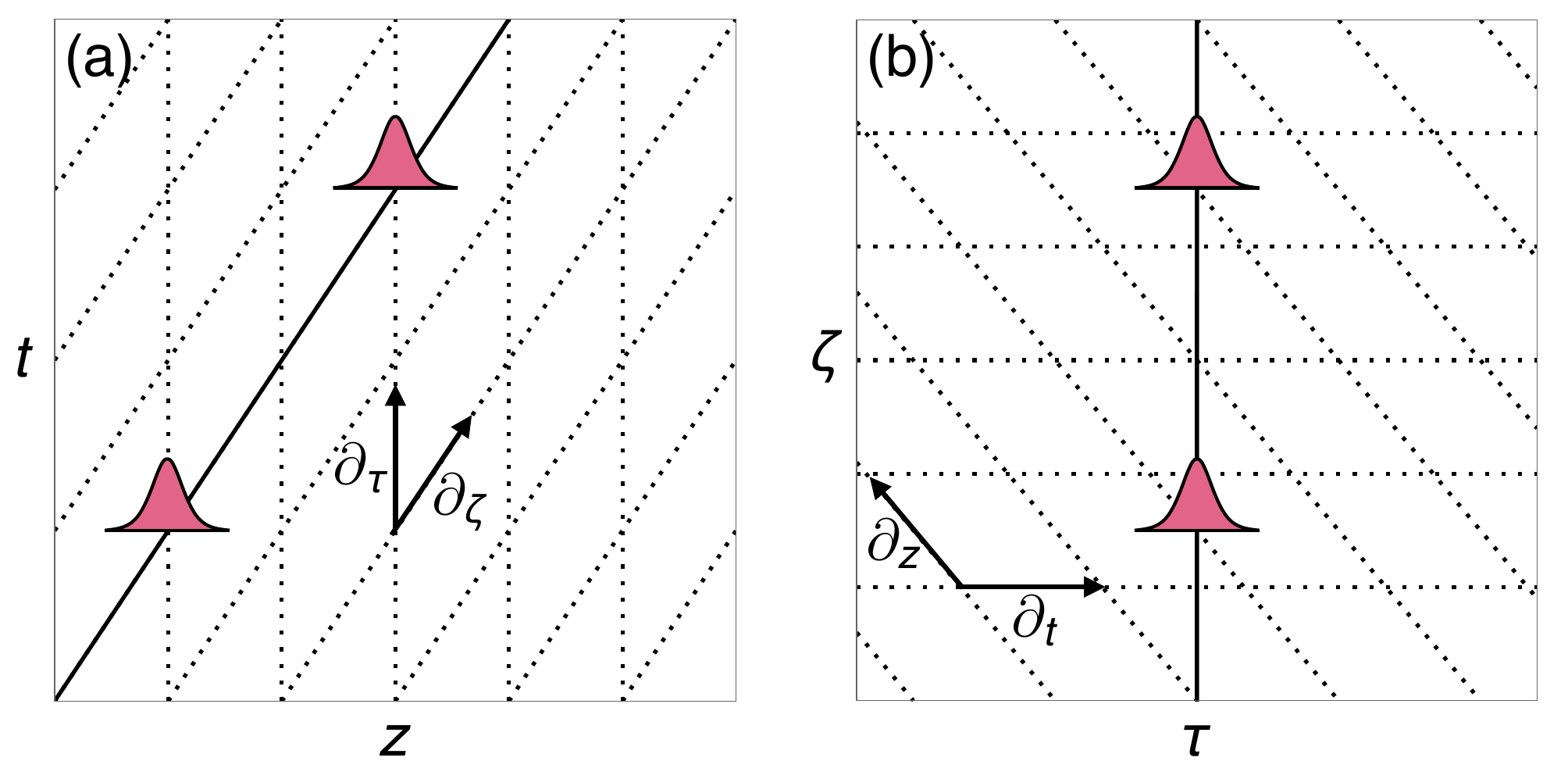}
	\caption{(a)An optical pulse traveling in the laboratory coordinates $(z,t)$. (b) An optical pulse traveling in the comoving (retarded) ($\zeta$,$\tau$) coordinates.}
	\label{fig.Fibers}
\end{figure}

Using the chain rule, the derivatives transform as
\begin{align}
	\frac{\partial}{\partial t}
	&= \frac{\partial \tau}{\partial t}\frac{\partial}{\partial \tau}+\frac{\partial \zeta}{\partial t}\frac{\partial}{\partial \zeta}=\frac{\partial}{\partial \tau}, \\
	\frac{\partial}{\partial x}&= \frac{\partial \tau}{\partial x}\frac{\partial}{\partial \tau}+\frac{\partial \zeta}{\partial x}\frac{\partial}{\partial \zeta}=\frac{1}{u}\left( \frac{\partial}{\partial \zeta}+\frac{\partial}{\partial \tau}\right).
\end{align}

There is a change in the refractive index due to the optical Kerr effect according to
\begin{equation}
	n(x,t) = n_0 + \delta n(x')=n_0 + \delta n(x,t)= n_0 + \delta n(\tau).
\end{equation}
Since the Kerr-induced change in the refractive index depends only on the retarded time, $\delta n = \delta n(\tau)$, its derivative in the comoving frame can be written as
\begin{equation}
	\frac{\dd n}{\dd x'} = \frac{\dd \,\delta n}{\dd x'}
	= \gamma\!\left(\frac{\partial}{\partial x}-\frac{u}{c^2}\frac{\partial}{\partial t}
	\right)\delta n .
\end{equation}

Expressing laboratory derivatives in terms of retarded coordinates
$(\tau,\chi)$,
\begin{equation}
	\frac{\partial}{\partial x}=\frac{\partial}{\partial \zeta}+\frac{1}{u}\frac{\partial}{\partial \tau},\qquad\frac{\partial}{\partial t} \frac{\partial}{\partial \tau},
\end{equation}
and using $\partial_{\zeta}\delta n = 0$, we obtain
\begin{equation}
	\frac{\dd n}{\dd x'}=\gamma\!\left(\frac{1}{u}-\frac{u}{c^2}\right)\frac{\partial \delta n}{\partial \tau} \frac{\gamma}{u}\!\left(1-\frac{u^2}{c^2}\right)\frac{\partial \delta n}{\partial \tau}=\frac{1}{\gamma u}\frac{\partial \delta n}{\partial \tau}.
\end{equation}

Returning to the definition of the surface gravity in Eq. \eqref{SGO} and using the horizon condition $u = -c/n$, we find
\begin{equation}
	\alpha'=\frac{c^2}{n^2}\frac{n}{\gamma c}\frac{\partial \delta n}{\partial\tau}\Big|_{h}=\frac{\gamma}{n}\frac{\partial \delta n}{\partial \tau}\Big|_{h}.
\end{equation}
Then, the surface gravity in the comoving frame is
\begin{equation}
	\alpha'=\frac{\gamma}{n}\frac{\partial \delta n}{\partial \tau}\Big|_{h}.
\end{equation}

\subsection{Doppler effect to the laboratory frame}\label{sec.Doppler}

There is a Doppler effect on the frequency when changing from the laboratory frame to the comoving frame. To show this, first we take the Planck spectrum
\begin{equation}
	\bar n=\frac{1}{z}\sum_{n=0}^{\infty}ne^{-n\beta}=-\frac{1}{z}\frac{\partial z}{\partial \beta}=\frac{1}{e^{\beta}-1},
\end{equation}
where
\begin{equation}
	\beta=\frac{\hbar\omega'}{k_B T},
\end{equation}
and $\omega'$ is the frequency in the comoving frame. The frequency transforms as
\begin{equation}
	\omega'=\gamma\omega\left(1+\frac{u}{c}n_0\right),
\end{equation}
$u$ is the velocity of the pulse and $n_0$ is the refractive index of the background. At the horizon, the velocity satisfies that $u=-c/n$, then the transformation between frames is 
\begin{equation}
	\omega'=\gamma\omega\left(1-\frac{n_0}{n}\right)=\gamma\omega\frac{n-n_0}{n}.
\end{equation}
Defining $\left.\delta n\right|_h\equiv n-n_0$, where $h$ is the horizon. In this case this is a certain $T_h$. We obtain that near the horizon
\begin{equation}
	\omega'=\omega\frac{\gamma}{n}\left.\delta n\right|_h.
\end{equation}
We have shown that $\alpha'$ scales with the same factor
\begin{equation}
	\alpha'=\alpha\frac{\gamma}{n}\left. \delta n\right|_h.
\end{equation}
Then, performing the inverse transformation 
\begin{equation}
	\alpha=\frac{n\alpha'}{\gamma\left.\delta n\right|_h}=\frac{n}{\gamma\left.\delta n\right|_h}\frac{\gamma}{n}\left.\frac{\dd\delta n}{\dd\tau}\right|_h=\left.\left(\frac{1}{\delta n}\frac{\dd\delta n}{\dd\tau}\right)\right|_h.
\end{equation}

This result implies that Hawking radiation exists in the laboratory frame and it depends on relative change of $\delta n$. This change can produce a measurable Hawking temperature $T_H$ if the variation in $\tau$ is ultrafast, that is, of the scale of a single optical cycle. This explains why we need to work in the extreme nonlinear optics (XNLO) regime.

\section{Pulse propagation in fibers}\label{sec.propagation}
\subsection{Pulse dynamics}\label{sec.dynamics} 

The study of light inside a dielectric material is governed by Maxwell's equations without free charges nor currents, i.e.,
\begin{alignat}{2}
	\nabla\cdot\mb{D}&=0,&\qquad \nabla\times\mb{E}&=-\frac{\partial\mb{B}}{\partial t},\\
	\nabla\cdot\mb{D}&=0,&\qquad \nabla\times\mb{E}&=\frac{\partial\mb{D}}{\partial t}.
\end{alignat}
Together with the constitutive equations
\begin{equation}
	\mb{D}=\epsilon_0\mb{E}+\mb{P},\qquad
	\mb{B}=\mu_0(\mb{H}+\mb{M}),
\end{equation}
it is possible to derive a wave equation for the electric field $\mb{E}$ as
\begin{equation}
	\nabla^2\mb{E}-\frac{1}{c^2}\frac{\partial^2\mb{E}}{\partial t^2}=\frac{1}{\epsilon_0c^2}\frac{\partial^2\mb{P}}{\partial t^2}
\end{equation}
where $\mb{P}$ is the polarization vector, which can be separated into linear and nonlinear parts for convenience $\mb{P}=\mb{P}_\text{L}+\mb{P}_\text{NL}$, and
\begin{equation}
	\mb{P}_\text{NL}=\epsilon_0\left(\chi^{(2)}:\mb{E}^2+\chi^{(3)}\vdots \mb{E}^3+\cdots\right).
\end{equation}

We can define the electric permittivity as $\epsilon(\omega)=1+\chi^{(1)}(\omega)$ and its relation with the refractive index $n^2(\omega)=\epsilon(\omega)$. In this way we can finally reach the wave equation known as the forward Maxwell equation (FME)
\begin{equation}
	\nabla^2 \mb{E}+\beta^2(\omega)\mb{E} +\frac{\omega^2}{\epsilon_0 c^2} \mb{P}_\text{NL}=0, \label{FME}
\end{equation}
where $\beta(\omega)=\omega^2 n^2(\omega)/c^2$. The three terms in Eq. \eqref{FME} describe the propagation, the dispersion and the nonlinearity, respectively. In the next two sections we will focus on the description of the dispersion and the nonlinearity.

\subsection{Dispersion}\label{sec.dispersion}
\begin{figure}
	\centering
	\includegraphics[width=0.6\linewidth]{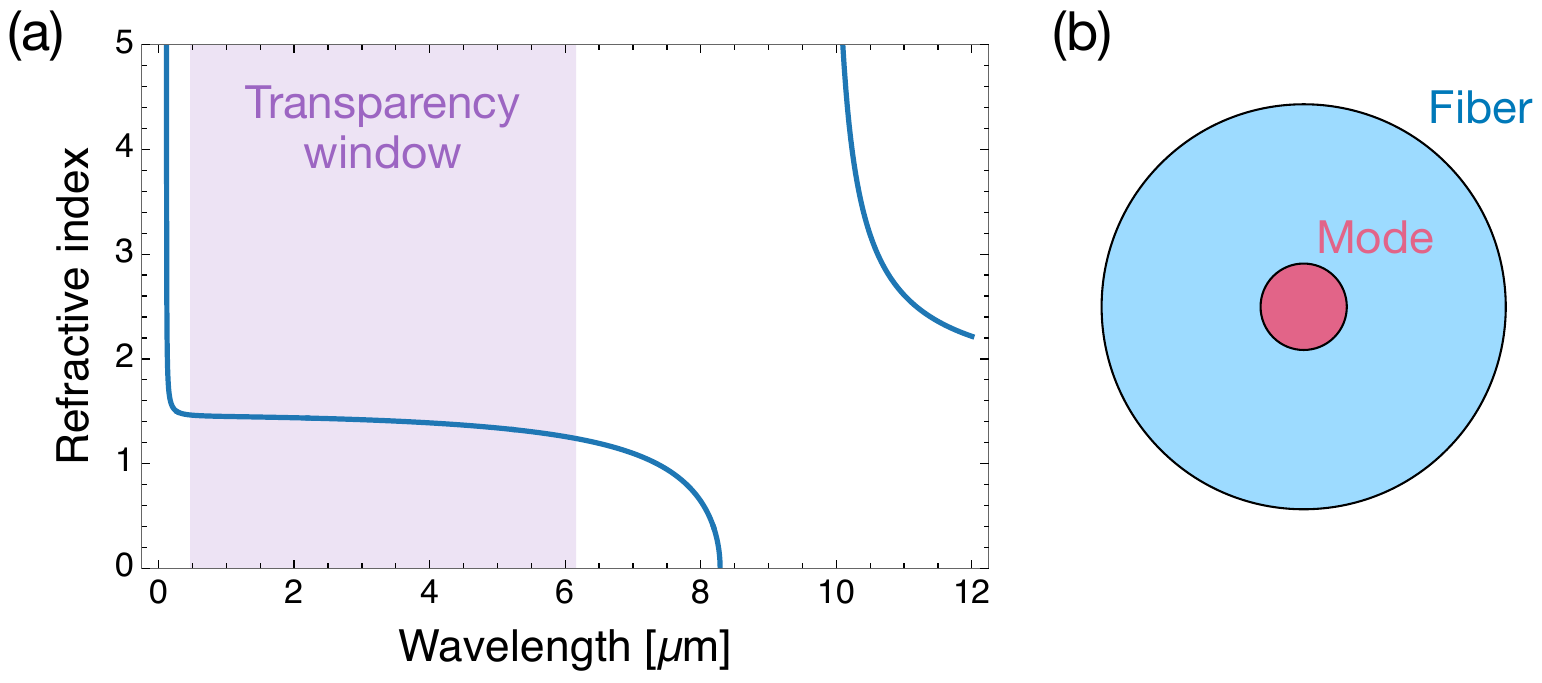}
	\caption{(a) Frequency dependence of the refractive index described by the Sellmeier model. Material dispersion originates from optical resonances, producing a smooth variation of the refractive index within the transparency window and rapid changes near resonant frequencies. The shaded region indicates the spectral range where absorption is weak and light propagates efficiently in the fiber. (b) Transversal plane of a fiber and its mode function.}
	\label{fig.Transparency}
\end{figure}

Dispersion in optical fibers arises from the frequency dependence of the refractive index. Material dispersion is given by its resonances and is captured by the Sellmeier equation.
\begin{equation}
	n^2(\omega)=1+\sum_j\frac{B_j\omega^2}{\omega_j^2-\omega^2},
\end{equation}
where the refractive index $n(\omega)$ varies smoothly within the transparency window, but changes rapidly near resonances $j$, see Fig. \ref{fig.Transparency}. This part of the dispersion is known for silica, the material of the PCFs.

The frequency dependence translates into the propagation constant $\beta$ to give a dispersive wavenumber function
\begin{equation}
	k(\omega)=\beta(\omega)=\frac{\omega}{c}n(\omega).
\end{equation}

This can be characterized using a Taylor expansion around a central frequency $\omega_0$
\begin{equation}
	\beta(\omega)=\sum_m\frac{\beta_m}{m!}(\omega-\omega_0)^m; \quad \beta_m=\frac{\dd^m\beta}{\dd \omega^m}.
\end{equation}
Each term in this expansion captures different dispersive behavior. The first order term $\beta_1$, relates to the group velocity as
\begin{equation}
	\beta_1=\frac{1}{v_g}=\frac{n_g}{c}=\frac{\dd}{\dd\omega}\left(\frac{\omega n(\omega)}{c}\right)=\frac{1}{c}\left(n+\omega\frac{\dd n}{\dd\omega}\right).
\end{equation}
From the above equation we can define a group index $n_g$ as
\begin{equation}
	n_g=n+\omega\frac{\dd n}{\dd\omega}.
\end{equation}

The second order term $\beta_2$ defines what is known as the group velocity dispersion (GVD)
\begin{equation}
	\beta_2=\frac{\dd^2}{\dd\omega^2}\left(\frac{\omega n}{c}\right)=\frac{1}{c}\frac{\dd}{\dd\omega}\left( n+\omega\frac{\dd n}{\dd\omega}\right)=\frac{1}{c}\left(2\frac{\dd n}{\dd\omega}+\omega\frac{\dd^2}{\dd\omega^2}\right).
\end{equation}
This term also allows us to define two different regimes: Normal GVD when $\beta_2>0$, where lower-frequency components travel faster than higher-frequency ones. Anomalous GVD when $\beta_2<0$, where higher-frequency components propagate with a larger group velocity than lower-frequency components. The anomalous GVD enables a range of nonlinear phenomena absent in the normal regime. 

\begin{figure}
	\centering
	\includegraphics[width=0.6\linewidth]{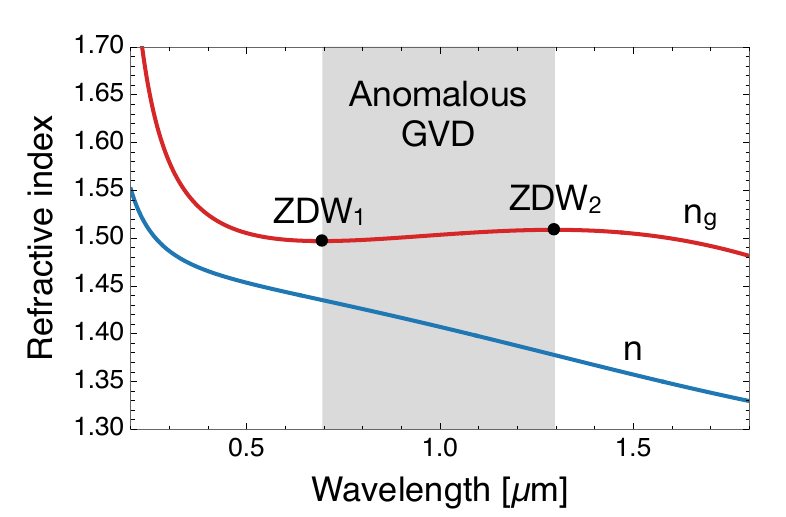}
	\caption{Refractive index $n$ and group index $n_g$ for a photonic crystal fiber. The two zero-dispersion wavelengths (ZDW$_1=700$ nm and ZDW$_2=1300$ nm) delimit the anomalous GVD region (shaded), where solitonic propagation is supported.}
	\label{fig.aGVD}
\end{figure}

To look at dispersive phenomena, let us consider the case of a photonic crystal fiber (PCF) made of silica. Its dispersion depends on the material and the geometry. The material dispersion is described by the Sellmeier resonances, which are known for silica. The geometry of the fiber depends on the air hole geometry. We define the zero-dispersion wavelengths (ZDWs) as the wavelengths $\lambda_\text{ZDW}$, where
\begin{equation}
	\beta_2(\lambda_{\text{ZDW}})=0.
\end{equation}
For a PCF similar to those used in the fiber-optical experiments of the analogue Hawking radiation \cite{Philbin.2008, Rubino.2012, Drori.2019, Felipe-Elizarraras.2022}: $\text{ZDW}_1=700$ nm and $\text{ZDW}_2=1300$ nm. In Fig. \ref{fig.aGVD} we show the refractive index $n$ and the group index $n_g$ in the visible and infrared range. We also work in the anomalous GVD region between the two ZDWs. In particular, the fiber supports solitonic solutions in this region \cite{Russell.2011}. 

\subsubsection{Fiber modes}
Let us consider a linearly polarized electric field $\mb{E}\Rightarrow E\hat{e}_z$. We can separate the longitudinal and transversal parts as
\begin{equation}
	\mb{E}(x,y,z)=E(z)F(x,y)\hat{e}_z.
\end{equation}
The equation for $F$ gives the effective propagation constant $\beta$.

\begin{figure}
	\centering
	\includegraphics[width=0.6\linewidth]{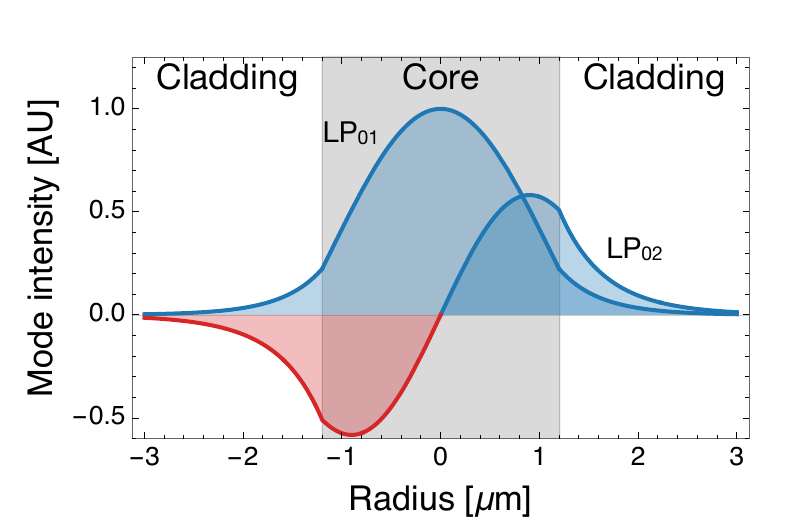}
	\caption{Transverse mode structure of a step-index optical fiber. The fundamental (LP$_{01}$) and first higher-order (LP$_{02}$) mode intensity profiles are shown as functions of the radial coordinate.}
	\label{fig.Modes}
\end{figure}

A standard model is the step-index fiber, in which the refractive index takes a higher constant value inside the core and drops abruptly in the cladding. The transverse mode equation admits solutions in terms of Bessel functions in the core region, which must be matched continuously to exponentially decaying fields in the cladding, see Fig. \ref{fig.Modes}.

In this regime, the modes are commonly classified as linearly polarized (LP) modes, denoted by $\text{LP}_{lm}$, where $l$ is the azimuthal order and $m$ the radial index. Although these modes are hybrid in nature, the LP approximation provides an accurate description. 

\begin{figure}
	\centering
	\includegraphics[width=0.6\linewidth]{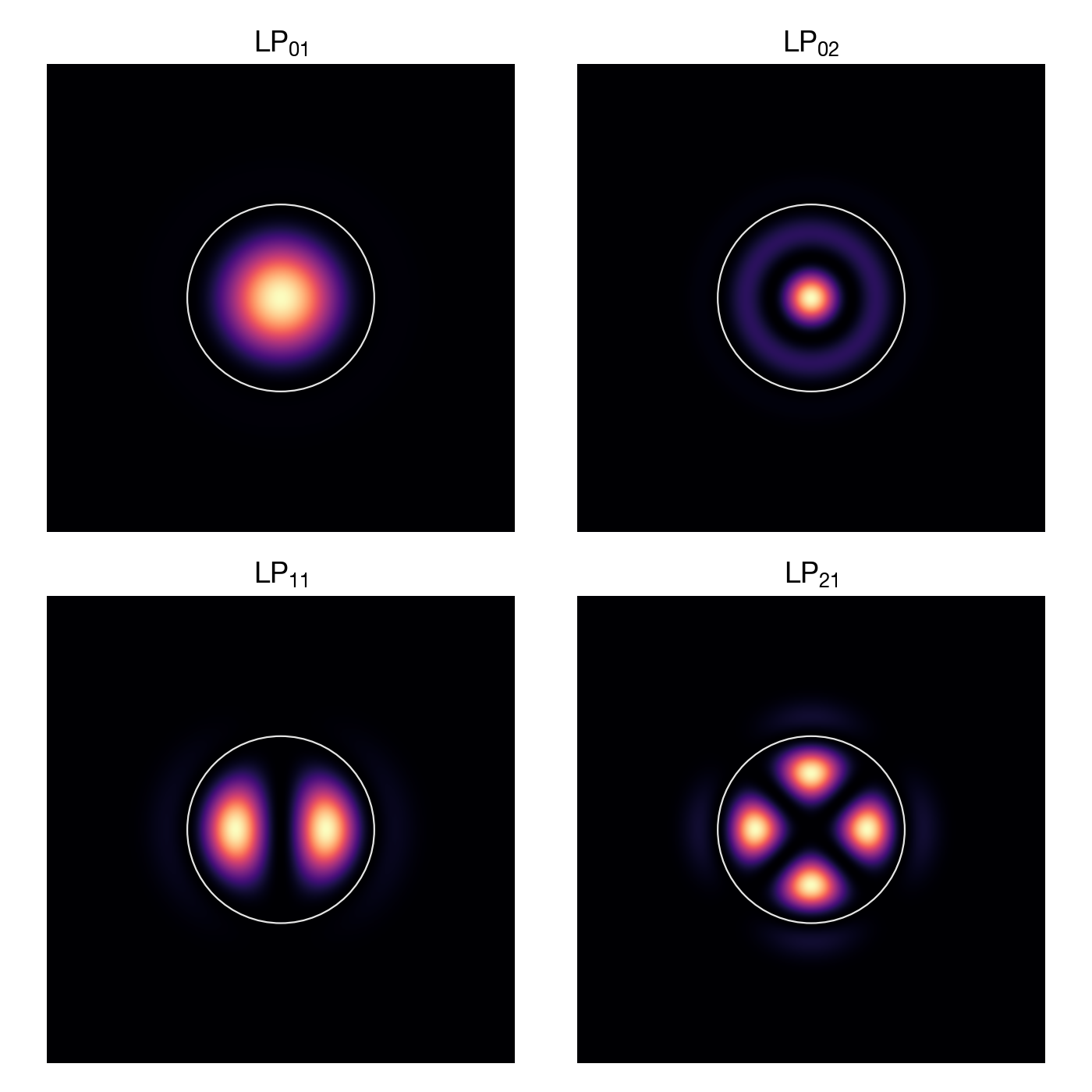}
	\caption{Transverse intensity profiles of representative linearly polarized modes in a step-index optical fiber.}
	\label{fig.LPModes}
\end{figure}

The fundamental mode $\text{LP}_{01}$ has a transverse intensity profile that is approximately Gaussian. Higher-order modes such as $\text{LP}_{02}$ and $\text{LP}_{03}$ or modes with nonzero angular structure $\text{LP}_{11}$ and $\text{LP}_{21}$ exhibit increasingly complex radial and azimuthal patterns, see Fig. \ref{fig.LPModes}. Each mode has its own dispersion relation $\beta(\omega)$, meaning that modal structure can influence the overall dispersive dynamics. We can assume excitation only in the fundamental mode, then the system reduces to a one-dimensional propagation problem along the $z$-axis.

\subsection{Nonlinearity}\label{sec.nonlinearity}
Taking a macroscopic approach to polarization as given by Bloembergen \cite{Bloembergen.1982}
\begin{equation}
	\mb{P}(\mb{r},t)=\mb{P}_{\text{L}}+\mb{P}_{\text{NL}}=\epsilon_0\sum_{j=1}^\infty \chi^{(j)}:\mb{E}^j=\epsilon_0(\chi^{(1)}\mb{E}+\chi^{(2)}:\mb{E}^2+\chi^{(3)}\vdots\mb{E}^3+...),
\end{equation}
where the linear part is just the $\chi^{(1)}$ term and the other terms are the nonlinear contribution. The term $\mb{P}_{\text{L}}$ is described by replacing $\epsilon_0\rightarrow\epsilon_0\epsilon$ and $\mu_0\rightarrow\mu_0\mu$. It is usual to keep only the first nonzero term for $\mb{P}_\text{NL}$. For example, a second-order nonlinearity, $\chi^{(2)}\neq 0$ occurs in nonlinear crystals, such as beta barium borate ($\beta$-BBO) and it produces effects such as the second harmonic generation (SHG), spontaneous parametric down conversion (SPDC), and the Pockels effect. 

When $\chi^{(2)}= 0$, the molecules of the material are symmetric, and therefore in general $\chi^{(3)}\neq 0$ so we have a third-order nonlinearity, which appears in optical fibers and produces the Kerr effect, self-phase modulation (SPM), spontaneous four-wave mixing (SFWM), third harmonic generation (THG), and, in particular, it is the interaction that produces analogue Hawking radiation.

For an electric field $\mb{E}$ copolarized in a single direction in a $\chi^{(3)}$ material
\begin{equation}
	\mb{P}=\epsilon_0\chi^{(3)}\vdots\mb{E}^3=\epsilon_0\chi^{(3)}E^3\hat{z}.
\end{equation}

Performing a change from the electric field to the analytic signal \cite{Amiranashvili.2016}:
\begin{align}
	&E\in\mathbb{R},\pm\omega,\qquad\mathcal{E}\in\mathbb{C},+\omega,\\
	&E=E_0\cos\theta=E_0\frac{e^{i\theta}+e^{-i\theta}}{2}=\frac{E_0e^{i\theta}+E_0e^{-i\theta}}{2}=\frac{\mathcal{E}+\mathcal{E}^*}{2}.\label{analysig}
\end{align} 
The analytic signal is defined with the Hilbert transform
\begin{equation}
	\mathcal{E}(z,t)=\frac{1}{\pi}\int_0^{\infty}\tilde{E}(z,\omega)e^{-i\omega t}\dd\omega.
\end{equation}

Returning to FME equation in Eq. \eqref{FME} with the analytic signal,
\begin{align}
	\nabla^2\mb{\mathcal{E}}+\beta^2(\omega)\mb{\mathcal{E}}+\frac{\omega^2}{\epsilon_0c^2}\mb{P}_{\text{NL}}=0,
\end{align}
Considering that the system is (1+1)D and the nonlinearity is given by the SPM term responsible for the Kerr effect, we have
\begin{align}
	\partial_z^2\mathcal{E}+\beta^2(\omega)\mathcal{E}+\frac{\omega^2}{c^2}\frac{3\chi^{(3)}}{4}|\mathcal{E}|^2\mathcal{E}=0, \\
	\mb{P}_{\text{NL}}=\frac{3}{4}\epsilon_0\chi^{(3)}|\mathcal{E}|^2\mathcal{E}.
\end{align}
We can again consider the QFTCS approximation of separating the field $\mathcal{E}$ as $\mathcal{E}=\mathcal{E}_0+\psi$, with $\mathcal{E}_0$ as the background and $\psi$ as the fluctuation. Then, considering linear terms in $\psi$, we keep the interaction term $|\mathcal{E}_0|^2\mathcal{E}$, where we have
\begin{align}
	\nabla^2\mathcal{E}+\left(\frac{\omega^2n^2}{c^2}+\frac{\omega^2}{c^2}\frac{3\chi^{(3)}}{4}|\mathcal{E}_0|^2\right)\mathcal{E}&=0,\\
	\partial_z^2\mathcal{E}+\frac{\omega^2}{c^2}\left(n^2+2n_0n_2I\right)\mathcal{E}&=\partial_z^2\mathcal{E}+\frac{\omega^2}{c^2}\left(n+n_2I\right)^2,\\
	\partial_z^2\mathcal{E}+\frac{\omega^2n_{\text{eff}}^2(\omega,\tau)}{c^2}\mathcal{E}&=0,
\end{align}
where we defined $2n_0n_2=3\chi^{(3)}/4$, then, $n_2=3\chi^{(3)}/8n_0$ and $n_{\text{eff}}(\omega,\tau)=n(\omega)+n_2I(\tau)$.

If $A$ satisfies a wave equation, all of its time derivatives do so. For the longitudinal components,
\begin{equation}
	\partial_z\tilde{A}+\beta^{2}(\omega)\tilde{A}+\omega^2\chi(\omega)\tilde{A}=0.
\end{equation}

Returning to the time domain with $\omega\rightarrow i\partial_t$, we get
\begin{equation}
	\partial_z^2\mathcal{E}+\beta^2(i\partial_t)\mathcal{E}-\frac{1}{\epsilon_0}\partial_t(\chi\partial_t)\mathcal{E}=0, \label{wetd}
\end{equation}
with $\chi=\chi^{(3)}|\epsilon_0|^2/c^2$.

Then, performing the coordinate transformation into the retarded frame,
\begin{equation*}
	\tau=t-\frac{z}{u},\qquad\zeta=\frac{z}{u},
\end{equation*}
which is noninertial but a valid reference frame. The derivatives transform as
\begin{align}
	\frac{\partial}{\partial t}&=\frac{\partial\tau}{\partial t}\frac{\partial}{\partial\tau}+\frac{\partial\zeta}{\partial t}\frac{\partial}{\partial\zeta}=\frac{\partial}{\partial\tau},\\
	\frac{\partial}{\partial z}&=\frac{\partial\tau}{\partial z}\frac{\partial}{\partial\tau}+\frac{\partial\zeta}{\partial z}\frac{\partial}{\partial\zeta}=\frac{1}{u}(\partial_\zeta-\partial_\tau).
\end{align}

Then $\beta(i\partial_t)\rightarrow -\beta_2\partial_\tau^2$, and substituting into Eq. \eqref{wetd}
\begin{align}
	\frac{1}{u^2}(\partial_\zeta-\partial_\tau)^2\mathcal{E}+\beta^2(i\partial_\tau)\mathcal{E}-\partial_\tau\left(\frac{\chi}{\epsilon_0}\partial_\tau\right)\mathcal{E}&=0,\\
	(\partial_\zeta-\partial_\tau)^2\mathcal{E}-\partial_\tau\left(\frac{u^2}{c^2}n^2\partial_\tau\right)\mathcal{E}&=0,\\
	\left[\partial_{\zeta}^2-\partial_\zeta\partial_\tau-\partial_\tau\partial_\zeta+\partial_\tau\left(1-\frac{u^2}{c^2}n^2\right)\partial_\tau\right]\mathcal{E}&=0.
\end{align}

Let us rewrite this last equation in relativistic form
\begin{equation}
	\partial_\mu g^{\mu\nu}\partial_\nu\mathcal{E}=0,
\end{equation}
where $g^{\mu\nu}$ is the inverse metric tensor and
\begin{equation}
	g^{\mu\nu}=\begin{pmatrix} 1&-1\\-1&1-\frac{u^2n^2}{c^2}\end{pmatrix}.
\end{equation}

To obtain the metric tensor we obtain the determinant 
\begin{equation}
	\frac{1}{g}=\det(g^{\mu\nu})=-\frac{u^2n^2}{c^2},
\end{equation}
finally
\begin{equation}
	g_{\mu\nu}=\displaystyle\frac{c^2}{u^2n^2}\begin{pmatrix} \frac{u^2n^2}{c^2}-1&-1\\-1&-1\end{pmatrix}.
\end{equation}

We can define the horizon from the metric tensor as $g_{00}=0$; then $(u^2n^2/c^2)-1=0$ implies that $v_p=\pm u$.

\subsection{Unidirectional approximation}
We separate the forward and backward propagating components of the field using the factorized Helmholtz equation.
\begin{equation}
	(\partial_z+i\beta(\omega))(\partial_z-i\beta(\omega))\tilde{\mathcal{E}}_\omega=-\frac{\omega^2}{\epsilon_0c^2}\tilde{\mb{P}},
\end{equation}
This equation has a general solution in terms of a superposition of forwards and backwards propagating modes
\begin{equation}
	\tilde{\mathcal{E}}(\omega,z)=\tilde{\mathcal{E}}_+(\omega)e^{i\beta z}+\tilde{\mathcal{E}}_-(\omega)e^{-i\beta(\omega)z}.
\end{equation}

If we assume that the backward-propagating modes are negligible, i.e., $|\tilde{\mathcal{E}}_-|\ll |\tilde{\mathcal{E}}_+|$, we can reduce the operator acting on the forward modes as
\begin{equation}
	\partial_z+i\beta(\omega)\simeq 2i\beta(\omega).
\end{equation}

Then the equation becomes
\begin{equation}
	i\partial_z\tilde{\mathcal{E}}_\omega+\beta(\omega)\tilde{\mathcal{E}}_\omega=i\frac{\omega^2}{2u^2\beta(\omega)}\frac{\tilde{\mb{P}}_\text{NL}}{\epsilon_0},
\end{equation}
where we do not write the subindex $+$.
Finally, we obtain the unidirectional pulse propagation equation (UPPE)
\begin{equation}
	i\partial_z\tilde{\mathcal{E}}_\omega+\beta(\omega)\tilde{\mathcal{E}}_\omega+\frac{\omega}{2cn(\omega)}\frac{\tilde{\mb{P}}_\text{NL}}{\epsilon_0}=0. \label{uppe}
\end{equation}

\section{Symmetries of the nonlinear Schrödinger equation}\label{sec.NLSElab}
Many complex physical phenomena are best modeled using nonlinear partial differential equations (NPDEs). One of the most widely studied NPDEs is the nonlinear Schr\"odinger equation (NLSE). This equation appears in various areas of physics, such as fluid mechanics, hydrodynamics, plasma physics, superconductivity, condensed matter physics, and nonlinear optics \cite{Fibich.2015}. An important example is in Bose-Einstein condensates where the NLSE is formally equivalent to the Gross--Pitaevskii equation \cite{Karjanto.2019}.

In this work, we focus on the context of nonlinear optics, where the NLSE is used to study pulse propagation in waveguides or fibers. It is the simplest equation that captures the interplay between dispersion and nonlinearity and, importantly, it supports solitonic solutions. Here, in particular, we study its symmetries and conserved quantities via Noether's theorem.

\subsection{NLSE in nonlinear optics}
As we have shown in previous sections, the dynamics of an optical pulse can be analyzed either in the laboratory $(t, z)$ or in the comoving frame $(\tau, \zeta)$. To derive the NLSE in the laboratory frame, we start from the unidirectional pulse propagation equation (UPPE) in Eq. \eqref{uppe}.

For a pulse propagating along the $z$-axis under the unidirectional approximation and assuming a narrowband pulse, the UPPE reduces to 
\begin{equation}
\partial_z \tilde{\mathcal{E}}(z, \omega) = i [\beta(\omega) - \beta(\omega_0)] \tilde{\mathcal{E}}(z, \omega) + i \frac{\omega^2}{2c^2\beta(\omega)} \tilde{P}_\text{NL}(z, \omega),
\end{equation}
where the wavenumber is defined by the dispersion relation $k^2(\omega) = \omega^2\epsilon(\omega)/c^2$ and $n^{2}(\omega)=\epsilon(\omega)$. Following the standard approach \cite{Agrawal.2013}, the effective dielectric constant  $\epsilon$ includes the nonlinear Kerr response of the medium,
\begin{equation}
	\epsilon(\omega) = 1 + \chi^{(1)}(\omega) + \frac{3}{4} \chi^{(3)} |\mathcal{E}(z,t)|^{2}.
\end{equation}
By performing a Taylor expansion of the propagation constant $\beta(\omega)$ around a central frequency $\omega_0$:
\begin{equation}
\beta(\omega) \approx \beta_0 + \beta_1(\omega-\omega_0) + \frac{1}{2}\beta_2(\omega-\omega_0)^2 + \dots,
\end{equation}
where $\beta_n = \left.\partial^n_\omega\beta\right|_{\omega_0}$. Substituting this expansion into the propagation equation and transforming back to the time domain via the substitution $i(\omega - \omega_0) \rightarrow \partial_t$, we obtain the NLSE in the laboratory frame:
\begin{equation}
i\partial_z \mathcal{E} + i\beta_1(\omega_0)\partial_t \mathcal{E}-\frac{1}{2}\beta_2(\omega_0)\partial_t^2 \mathcal{E}= 0.
\end{equation}

To simplify the analysis, we switch to the comoving frame defined by the coordinates in Eq. \eqref{coco}, where the flow velocity is $u = 1/\beta_1$. The NLSE in the comoving frame for the analytic signal is
\begin{equation}
i\partial_\zeta \mathcal{E} - \frac{1}{2} \beta_2 \partial_\tau^2 \mathcal{E}+ \gamma \left|\mathcal{E}\right|^{2}\mathcal{E}= 0,\label{NLSEco}
\end{equation}
where the nonlinear term $\left|\mathcal{E}\right|^{2}\mathcal{E}$ represents self-phase modulation (SPM) \cite{Agrawal.2013}. The nonlinear coefficient $\gamma$ is given by the overlap integral of the fiber mode $B(x,y)$ as
\begin{equation}
	\gamma = \gamma(\omega_0) = \frac{n_2 \omega_0}{c A_\text{eff}}, \qquad
	A_\text{eff} = \frac{\left(\iint |B(x, y)|^2 \, \mathrm{d}x \, \mathrm{d}y\right)^2}{\iint |B(x, y)|^4 \, \mathrm{d}x \, \mathrm{d}y}.
\end{equation}

\subsection{Lagrangian formalism of the NLSE}
To establish the Lagrangian formalism for the NLSE, we approach the problem from the inverse perspective of classical field theory: while the equation of motion is known, the corresponding Lagrangian density must be constructed. We perform this analysis in the laboratory frame $(t, z)$ to facilitate the interpretation of the conserved quantities before returning to the comoving frame $(\zeta,\tau)$.

We consider a version of the NLSE with anomalous GVD $\beta_2<0$ and repulsive nonlinearity $\gamma>0$.
\begin{equation}
i\partial_{t}\mathcal{E}+\frac{\beta_{2}}{2}\partial_{z}^{2}\mathcal{E}+\gamma \left|\mathcal{E}\right|^{2}\mathcal{E}=0.    
\end{equation}
By examining the structure of the NLSE, we propose the following Lagrangian density \cite{Kudryashov.2023}
\begin{equation}
	\mathcal{L} = \frac{i}{2} \left( \mathcal{E}^* \partial_{t} \mathcal{E} - \mathcal{E} \partial_{t} \mathcal{E}^* \right) - \frac{\beta_{2}}{2} \left| \partial_{z} \mathcal{E} \right|^{2} + \frac{\gamma}{2} \mathcal{E}^{2}\mathcal{E}^{*2}. \label{NLSEL}
\end{equation}
Since the action is a functional of independent variables, we treat the fields $\mathcal{E}$ and $\mathcal{E}^{*}$ as independent within the calculus of variations. Applying the Euler--Lagrange equations
\begin{equation}\label{E-L}
\frac{\delta L}{\delta \phi_{\alpha}} = \frac{\partial \mathcal{L}}{\partial \phi_{\alpha}} - \partial_{\beta} \frac{\partial \mathcal{L}}{\partial (\partial_{\beta} \phi_{\alpha})}= 0,
\end{equation}
where $\phi_\alpha\in \{\mathcal{E}, \mathcal{E}^*\}$ represent the set of independent canonical fields. For the field $\mathcal{E}$, the variation yields
\begin{align}
\frac{\partial \mathcal{L}}{\partial \mathcal{E}^{*}}-\partial_{\mu}\frac{\partial \mathcal{L}}{\partial(\partial_{\mu}\mathcal{E}^{*})}=\frac{\partial \mathcal{L}}{\partial \mathcal{E}^{*}}-\partial_{t}\frac{\partial \mathcal{L}}{\partial(\partial_{t}\mathcal{E}^{*})}-\partial_{z}\frac{\partial \mathcal{L}}{\partial(\partial_{z}\mathcal{E}^{*})} &= 0 
\\
i\partial_{t}\mathcal{E}+\frac{\beta_{2}}{2}\partial_{z}^{2}\mathcal{E}+\gamma\left|\mathcal{E}\right|^{2}\mathcal{E} &=0,
\end{align}
where we utilize the summation over indices $\mu=0,1$, corresponding to coordinates $t$ and $z$. An analogous procedure for $\mathcal{E}^*$ recovers the complex conjugate equation. This verification confirms that Eq. \eqref{NLSEL} is the correct Lagrangian density for the NLSE.

\subsection{Conserved charges}
We now turn to the analysis of the symmetries of the NLSE. By Noether's first theorem, each continuous symmetry corresponds to a conservation law expressed as a continuity equation
\begin{equation}\label{ConsL}
\partial_{t}\rho(z,t)+\partial_{z}j(z,t) = 0,     
\end{equation}
where $\rho$ is the charge density and $j$ is the density flux. The corresponding conserved quantity is the total charge $Q$, given by
\begin{equation}
Q=\int_{-\infty}^{\infty}\rho(z,t) \,\text{d}z.    
\end{equation}
In the context of optical fibers, we apply this framework to three fundamental symmetries.

\subsubsection{U(1) symmetry}
We first verify that the Lagrangian of the NLSE in Eq. \eqref{NLSEL} is invariant under the $U(1)$ transformation $\mathcal{E}'=\mathcal{E}e^{i\varphi}$: 
\begin{align}
\mathcal{L}'&=\frac{i}{2}[\mathcal{E}'{}^{*}\partial_t(\mathcal{E}')-\mathcal{E}'\partial_t(\mathcal{E}'{}^{*})]
-\frac{\beta_2}{2}|\partial_z(\mathcal{E}')|^2+\frac{\gamma}{2}(\mathcal{E}'{}^2)(\mathcal{E}'{}^{*2})
\nonumber\\
&= \frac{i}{2}\left[\mathcal{E}^{*}e^{-i\varphi}\partial_{t}(\mathcal{E}e^{i\varphi})-\mathcal{E}e^{i\varphi}\partial_{t}(\mathcal{E}^{*}e^{-i\varphi})\right]-\frac{\beta_{2}}{2} \left|\partial_{z}(\mathcal{E}e^{i\varphi})\right|^{2}+\frac{\gamma}{2}(\mathcal{E}^{2}e^{2i\varphi})(\mathcal{E}^{*2}e^{-2i\varphi}) \nonumber\\
&= \frac{i}{2}\left(\mathcal{E}^{*}\partial_{t}\mathcal{E}-\mathcal{E}\partial_{t}\mathcal{E}^{*}\right)-\beta_{2}\left|\partial_{z}\mathcal{E}\right|^{2}+\frac{\gamma}{2}\mathcal{E}^{2}\mathcal{E}^{*2}= \mathcal{L}.
\end{align}

For an infinitesimal transformation with $\varphi \to \delta \varphi$, the field variation is $\delta \mathcal{E} = i \delta \varphi \mathcal{E}$ or
\begin{equation}
\frac{\delta \mathcal{E}}{\delta \varphi} = i \mathcal{E}.
\end{equation}
Applying Noether's theorem, we obtain the equation
\begin{equation} 
\partial_{t}|\mathcal{E}|^{2}-i\frac{\beta_{2}}{2}\partial_{z}(\mathcal{E}^{*}\partial_{z}\mathcal{E}-\mathcal{E}\partial_{z}\mathcal{E}^{*})=0.
\end{equation}
This result matches the continuity equation form \eqref{ConsL}, allowing us to identify the conserved quantity as the total norm (or quasiparticle number):
\begin{equation}\label{nlse u(1)}
N =\int_{-\infty}^{\infty}|\mathcal{E}|^{2}\,\text{d}z.  
\end{equation}
In the literature, $\mathcal{N}= \left|\mathcal{E}\right|^{2}$ is typically interpreted as the photon-number density \cite{Fibich.2015} or its classical analogue.

\subsubsection{Spatial symmetry and linear momentum}
The action 
\begin{equation}
S = \int_{-\infty}^{\infty} \mathcal{L}\, \text{d}z\,\text{d}t = \int_{-\infty}^{\infty}\left[  \frac{i}{2}\left(\mathcal{E}^{*}\partial_{t}\mathcal{E}-\mathcal{E}\partial_{t}\mathcal{E}^{*}\right)-\frac{\beta_{2}}{2} \left|\partial_{z}\mathcal{E}\right|^{2}+\frac{\gamma}{2}\mathcal{E}^{2}\mathcal{E}^{*2} \right]\text{d}z\,\text{d}t,\label{action}
\end{equation}
is invariant under the spatial translation $z \to z + \delta z$. This continuous symmetry implies the existence of a conserved quantity related to the stress-energy tensor $(j_z, \rho_z)$ in the form of a continuity equation.

For the spatial shift, the conservation law takes the form
\begin{equation}
\frac{i}{2}\partial_{t}(\mathcal{E}^{*}\partial_{z}\mathcal{E}-\mathcal{E}\partial_{z}\mathcal{E}^{*})-\frac{1}{2}\partial_{z}\left[i\left(\mathcal{E}^{*}\partial_{t}\mathcal{E}-\mathcal{E}\partial_{t}\mathcal{E}^{*}\right)+\beta_{2}|\partial_{z}\mathcal{E}|^{2}+\gamma\mathcal{E}^{2}\mathcal{E}^{*2}\right]=0.
\end{equation}
This equation already has the form of a conservation equation. By integrating the density $\rho_z$ over the spatial domain, we find the conserved quantity, which is the linear momentum,
\begin{equation}\label{nlse momen}
P = \frac{i}{2}\int_{-\infty}^{\infty}\left(\mathcal{E}^{*}\partial_{z}\mathcal{E}-\mathcal{E}\partial_{z}\mathcal{E}^{*}\right)\,\text{d}z.
\end{equation}

\subsubsection{Temporal symmetry and the Hamiltonian}
The action in Eq. \eqref{action} is also invariant under a translation in time $t\rightarrow t+\delta t$. According to Noether's theorem, this symmetry leads to the conservation of the Hamiltonian, which in the context of nonlinear optics represents the total energy (or a quantity proportional to it) of the optical field.

By applying the transformation to the Lagrangian density $\mathcal{L}$, we define the Hamiltonian density $\mathcal{H}$ through the Legendre transformation of $\mathcal{L}$ with respect to the "velocity" $\partial_t\mathcal{E}$:
\begin{equation}
	\mathcal{H} = \frac{\partial \mathcal{L}}{\partial (\partial_t \mathcal{E})} \partial_t \mathcal{E} + \frac{\partial \mathcal{L}}{\partial (\partial_t \mathcal{E}^*)} \partial_t \mathcal{E}^* - \mathcal{L} .
\end{equation}
Substituting our corrected Lagrangian density into this expression, the terms involving time derivatives cancel out, leaving the spatial and nonlinear terms
\begin{equation}
	\partial_{t}\left(\frac{\beta_{2}}{2}|\partial_{z}\mathcal{E}|^{2}-\frac{\gamma}{2}\mathcal{E}^{2}\mathcal{E}^{*2}\right)-\frac{\beta_2}{2}\partial_{z}\left(\partial_{z}\mathcal{E}^{*}\partial_{t}\mathcal{E}+\partial_{z}\mathcal{E}\partial_{t}\mathcal{E}^{*}\right)=0.
\end{equation}
This equation already has the form of a continuity equation. We can identify the Hamiltonian density as
\begin{equation}
	\mathcal{H} = \frac{\beta_2}{2} \left| \partial_{z} \mathcal{E} \right|^{2} - \frac{\gamma}{2} \mathcal{E}^{2}\mathcal{E}^{*2}.
\end{equation}
By integrating the density $\mathcal{H}$ over the entire spatial domain, we obtain the conserved Hamiltonian $H$:
\begin{equation}\label{nlse hamiltonian}
H = \int_{-\infty}^{\infty} \left[    \beta_{2}|\partial_{z}\mathcal{E}|^{2}-\gamma\mathcal{E}^{2}\mathcal{E}^{*2}\right]\,\text{d}z,
\end{equation}
which remains constant during the evolution of the field $\mathcal{E}$ as long as the action remains explicitly independent of time. In physical terms, the first term represents the kinetic energy associated with the dispersion (GVD), while the second term represents the potential energy arising from the nonlinearity of the field (SPM). In the anomalous dispersion regime ($\beta_2<0$), these two terms can balance each other perfectly, giving rise to stable, non-diffracting wave packets known as solitons.

\subsection{Hamiltonian formalism of the NLSE}
The Hamiltonian $H$ was already obtained from Noether's theorem in Eq. \eqref{nlse hamiltonian} as the conserved quantity associated with temporal symmetry. However, it is instructive to rederive it via a Legendre transformation. 

Based on our Lagrangian in Eq. \eqref{NLSEL}, the canonical momenta are defined as
\begin{equation}
\pi \equiv \frac{\partial \mathcal{L}}{\partial (\partial_{t}\mathcal{E})} = \frac{i}{2}\mathcal{E}^{*}, \qquad 
\pi^{*} \equiv \frac{\partial \mathcal{L}}{\partial (\partial_{t}\mathcal{E}^{*})} = -\frac{i}{2}\mathcal{E}.      
\end{equation}
Consequently, the Hamiltonian density is
\begin{equation}
\mathcal{H}=\pi\,\partial_{t}\mathcal{E}+\pi^{*}\,\partial_{t}\mathcal{E}^{*} -\mathcal{L}=\frac{\beta_2}{2}|\partial_{z}\mathcal{E}|^{2}-\frac{\gamma}{2}\mathcal{E}^{2}\mathcal{E}^{*2},   
\end{equation}
which is identical to the form obtained previously from Noether's theorem.

We can also recover the NLSE from $\mathcal{H}$ by applying the complex version of Hamilton's equations \cite{Amiranashvili.2016}:
\begin{equation}\label{complexH}
i\partial_{t}a = \frac{\delta \mathcal{H}}{\delta a^{*} }= \frac{\partial \mathcal{H}}{\partial a^{*}}-\partial_{\beta}\frac{\partial \mathcal{H}}{\partial(\partial_{\beta} a^{*})}.    
\end{equation}
In our specific case, this yields
\begin{align}
 i\partial_{t}\mathcal{E} &= \frac{\delta \mathcal{H}}{\delta \mathcal{E}^{*} } = \frac{\partial \mathcal{H}}{\partial \mathcal{E}^{*}}-\partial_{\beta}\frac{\partial \mathcal{H}}{\partial(\partial_{\beta} \mathcal{E}^{*})}   \\ \nonumber
 i\partial_{t}\mathcal{E}&=-\frac{\beta_{2}}{2}\partial_{z}^{2}\mathcal{E}-\gamma\left|\mathcal{E}\right|^{2}\mathcal{E},
\end{align}
which is the NLSE.

Taking the complex conjugate of Eq. \eqref{complexH} gives the complex-conjugated NLSE. A conserved charge $Q$ can be obtained from the Hamiltonian if its Poisson bracket satisfies
\begin{equation}
\{Q, H\} = 0.
\end{equation}
This method is particularly valuable, as not all conserved quantities originate from continuous symmetries and thus from Noether's theorem. 

As an illustration, we show that $N$, $P$, and $H$ are indeed conserved quantities. For the photon number $N$:
\begin{align}
\{N, H\}&= \int_{-\infty}^{\infty}  \left[ \frac{\delta N}{\delta \mathcal{E}} \dot{\mathcal{E}}+ \frac{\delta N}{\delta \mathcal{E}^{*}} \dot{\mathcal{E}}^{*}  \right]\text{d}z
=\int_{-\infty}^{\infty}(\mathcal{E}^{*}\dot{\mathcal{E}}+\mathcal{E}\dot{\mathcal{E}}^{*})\text{d}z   \\ \nonumber  
&= i\frac{\beta_2}{2}\int_{-\infty}^{\infty}\partial_{z}(\mathcal{E}^{*}\partial_{z}\mathcal{E}-\mathcal{E}\partial_{z}\mathcal{E}^{*}) \,\text{d}z = 0.
\end{align}
Similarly, for the momentum $P$:
\begin{align}
\{P, H\}&= \int_{-\infty}^{\infty} \left[ \frac{\delta P}{\delta \mathcal{E}} \dot{\mathcal{E}}+ \frac{\delta P}{\delta \mathcal{E}^{*}} \dot{\mathcal{E}}^{*}  \right] \text{d}z =\int_{-\infty}^{\infty}[(-\partial_{z}\mathcal{E}^{*})\dot{\mathcal{E}}+(\partial_{z}\mathcal{E})\dot{\mathcal{E}}^{*}]\text{d}z    \\ \nonumber 
&=\int_{-\infty}^{\infty}\partial_{z}\left[\frac{\beta_2}{2}|\partial_{z}\mathcal{E}|^{2}+\frac{\gamma}{2}\mathcal{E}^{2}\mathcal{E}^{*2}\right]\text{d}z=0.
\end{align}
On the other hand, $\{H,H\}$ vanishes by construction if $H$ does not depend explicitly on $t$, as in our case
\begin{equation}
\{H, H\}= \int_{-\infty}^{\infty} \left[ \frac{\delta H}{\delta \mathcal{E}} \dot{\mathcal{E}}+\frac{\delta H}{\delta \mathcal{E}^{*}} \dot{\mathcal{E}}^{*}\right] \text{d}z =0.   
\end{equation}
Note that, due to the structure of the complex Hamilton's equations, the Poisson bracket $\{Q, H\}$ does not include the variational derivative with respect to $\pi$, since the canonical momentum is proportional to the field itself, making the derivative equivalent. 

Upon obtaining the Hamiltonian $H$, we can proceed to quantize the NLSE.  Under canonical quantization, the commutation relation
\begin{equation}
[\hat{\mathcal{E}}(z,t),\hat{\pi}(z',t)] = \delta(z-z'),
\end{equation}
with $\hat{\pi}=\hat{\mathcal{E}}^\dag$. This leads to the Hamiltonian operator
\begin{equation}
\hat{H}= \int_{-\infty}^{\infty} \left[\frac{\beta_2}{2}(\partial_{z}\hat{\mathcal{E}}^{\dagger})(\partial_{z}\hat{\mathcal{E}})
-\frac{\gamma}{2} \,\hat{\mathcal{E}}^{\dagger}\hat{\mathcal{E}}^{\dagger}\hat{\mathcal{E}}\hat{\mathcal{E}}\,\right]\text{d}z,
\end{equation}
where we assume normal ordering for simplicity, though the order of the nonlinear term can be derived from a microscopic model \cite{Drummond.2014} . Remarkably, this Hamiltonian shares the same mathematical structure as the Gross--Pitaevskii equation in Bose--Einstein condensates, despite the very different physical contexts. This structural similarity illustrates the fundamental importance of the NLSE across different physical domains.

\subsection{The nonlinear Schrödinger equation in the comoving frame}\label{sec.NLSEcm}
Now let us turn to the study of the NLSE in the comoving frame shown in Eq. \eqref{NLSEco}. This equation becomes particularly interesting when we analyze its symmetries in this reference frame, since they differ from those in the laboratory frame.

\subsubsection{Comoving frequency}
The motivation for changing reference frames arises from the interest in studying scenarios where an intense pump pulse defines a stationary background, allowing us to describe nonlinear optical processes in which a light pulse remains stationary in that frame. 

Let us consider a pump pulse in the soliton regime, denoted by $\mathcal{E}_0$, which establishes a background field through its interaction with the fiber via the Kerr effect. Superposed on this background is a fluctuation $\psi$ acting as a probe signal that evolves within the effective background defined by the pump. The background is stationary in the comoving frame, i.e., $\mathcal{E}_{0} = \mathcal{E}_{0}(\tau)\neq\mathcal{E}_{0}(\zeta)$.

To analyze the dynamics of the system, we perform a pump–probe separation $\mathcal{E} \rightarrow \mathcal{E}_{0} + \psi$ and substitute this expression into the NLSE in the comoving coordinates
\begin{equation}
i\partial_{\zeta}(\mathcal{E}_{0} + \psi) + \frac{\beta_{2}}{2}\partial_{\tau}^{2}(\mathcal{E}_{0} + \psi)+ \gamma|\mathcal{E}_0+\psi|^2(\mathcal{E}_0+\psi)  = 0.   
\end{equation}
Mathematically, this transformation introduces a perturbation into the field, analogous to the Bogoliubov--de Gennes formulation in Bose--Einstein condensates \cite{faccio.2013}. Expanding to first order in $\psi$, we obtain the linearized equation
\begin{equation}
i\partial_{\zeta}\mathcal{E}_{0} + \frac{\beta_{2}}{2}\partial_{\tau}^{2}\mathcal{E}_{0}+\gamma|\mathcal{E}_0|^2\mathcal{E}_0+ i\partial_{\zeta}\psi+ \frac{\beta_{2}}{2}\partial_{\tau}^{2}\psi+ \gamma [2\psi|\mathcal{E}_0|^2+\mathcal{E}_0^2\psi^{*}] = 0.
\end{equation}

The first three terms correspond to the NLSE for $\mathcal{E}_{0}$, which we assume is fulfilled, so we set them to zero. The remaining equation contains two nonlinear terms representing cross-phase modulation (XPM) and four-wave mixing (FWM). We focus explicitly on the XPM process given by the $\psi\psi^*$ term by applying the so-called rotating-wave approximation (RWA). Neglecting the counter-rotating components (i.e., dropping the $\psi^{*}$ terms), is justified under the counter-rotating wave approximation \cite{Drummond.2014}, since the FWM contribution is very weak due to phase-matching mismatch. This yields the evolution equation for the fluctuation as
\begin{equation}\label{eqpsi}
i\partial_{\zeta}\psi +\frac{\beta_2}{2}\partial_\tau^2 \psi+2\gamma|\mathcal{E}_0|^2\psi=0. 
\end{equation}
Note the factor of 2, which arises from the distinction between self-interaction and the interaction with a distinct background field (XPM). This equation can be derived from the Lagrangian density
\begin{equation}
\mathcal{L} = \frac{i}{2}\left(\psi^{*}\partial_{\zeta}\psi
- \psi\,\partial_{\zeta}\psi^{*}\right)
- \frac{\beta_{2}}{2}\lvert \partial_{\tau}\psi \rvert^{2}
+ \gamma (\mathcal{E}\mathcal{E}^{*} \psi\psi^{*}). \label{reduced_L}
\end{equation}

Since the background field $\mathcal{E}_0$ depends only on $\tau$ and not on $\zeta$, the Lagrangian density has no explicit dependence on $\zeta$. Therefore, it is invariant under translations $\zeta \rightarrow \zeta + \dd\epsilon$, and by Noether's theorem there exists a conserved quantity associated with this symmetry.

To determine it explicitly, we compute the canonical momenta conjugate to $\psi$ and $\psi^*$:
\begin{equation}
\pi_\psi=\frac{\partial \mathcal{L}}{\partial (\partial_\zeta \psi)}=\frac{i}{2}\psi^*,
\qquad
\pi_{\psi^*}=\frac{\partial \mathcal{L}}{\partial (\partial_\zeta \psi^*)}=-\frac{i}{2}\psi .
\end{equation}

The Hamiltonian density is defined as
\begin{equation}
\mathcal{H}=\pi_\psi \partial_\zeta \psi+\pi_{\psi^*} \partial_\zeta \psi^*-\mathcal{L},
\end{equation}
substituting the canonical momenta and the explicit form of $\mathcal{L}$ in Eq. \eqref{reduced_L}, the $\zeta$-derivative terms cancel identically, and we obtain
\begin{equation}
\mathcal{H}=\frac{i}{2}\psi^*\partial_\zeta\psi-\frac{i}{2}\psi\partial_\zeta\psi^*-\mathcal{L},
\end{equation}

Then, using the explicit form of $\mathcal{L}$ in Eq. \eqref{reduced_L}, the $\zeta$-derivative
terms cancel identically, and we obtain
\begin{equation}
\mathcal{H}=\frac{\beta_2}{2}|\partial_\tau \psi|^2-\gamma |\mathcal{E}_0(\tau)|^2 |\psi|^2.
\end{equation}

The corresponding conserved quantity is
\begin{equation}
H = \int d\tau \, \mathcal{H},
\end{equation}
which satisfies
\begin{equation}
\frac{dH}{d\zeta} = 0,
\end{equation}
Therefore,  modes of the form $\psi(\zeta,\tau) \propto e^{-i\omega' \zeta}$ propagate with constant comoving frequency $\omega'$. The conservation of $\omega'$ is a direct consequence of the translation invariance along $\zeta$.

However, in laboratory coordinates $(t,z)$ the background soliton is $\mathcal{E}_0(\tau) = \mathcal{E}_0(t - z/u)$, which depends explicitly on the laboratory time $t$ and propagation distance $z$. The Lagrangian therefore acquires explicit time and spatial dependence, breaking time-evolution and spatial translation symmetries in the laboratory frame. Consequently, neither the laboratory frequency $\omega$ nor the laboratory momentum $k$ are conserved individually; only a precise combination of them (the comoving frequency) remains invariant, defined by the velocity $u$.

\section{Timeline of fiber-optical experiments on analogue Hawking radiation}\label{sec.timeline}
A comprehensive history of analogue gravity itself can be found in the excellent review by Barceló, Liberati, and Visser \cite{Barcelo.2011}. This section gives a brief overview of this history, to then focus on a historical review of the timeline of the particular case of the fiber-optical experiments on the analogue Hawking radiation. The account is based on key experiment and they way they shape our understanding of the analogy and the nonlinear optics behind it.

Following the proposal of astrophysical Hawking radiation in 1974 \cite{Hawking.1974}, William Unruh proposed a similar effect in a moving fluid in 1981 \cite{Unruh.1981}. Although Unruh's work remained largely ignored for a decade, a seminal paper by Ted Jacobson in 1991 \cite{Jacobson.1991} rekindled scientific interest in the field. This sparked an intensive search for experimental implementations of analogue metrics. Researchers initially focused on superfluids---in particular Helium--3 and Helium--4 explored by Grigory Volovik \cite{Volovik.2003}---before expanding to other systems. These included water tanks \cite{Rousseaux.2008,Weinfurtner.2011}, Bose-Einstein condensates \cite{Garay.2000,Carusotto.2006} and fiber optics \cite{Philbin.2008}. This last system constitutes the focus of these notes.
		
\begin{figure}
\centering
\includegraphics[width=0.6\linewidth]{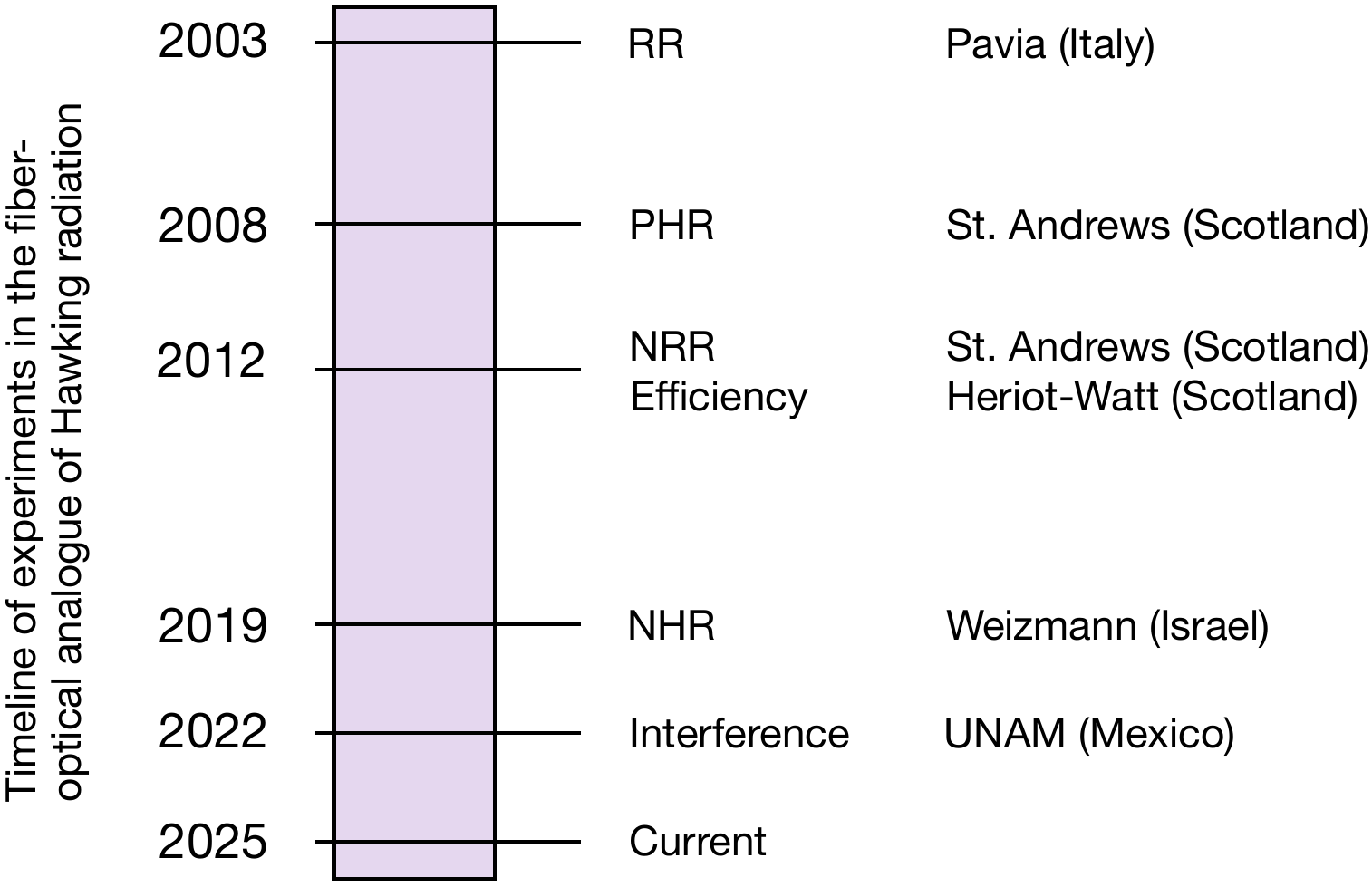}
\caption{Timeline of the fiber-experiments on analogue Hawking radiation and other resonant processes.}
\label{fig.timeline}
\end{figure}

The following sections detail each of the milestone experiments of analogue Hawking radiation in fiber optics, elucidating the relevant theory behind them. As this is an overview rather than a review, we highlight the pivotal developments while trying to acknowledge the vast body of work contributed by the broader scientific community. We apologize to all the scientists whose work is not included here.
		
\subsection{The 2008 St. Andrews experiment}\label{sec.2008}
After a failed attempt by Ulf Leonhardt in 2002 \cite{Leonhardt.2002} using slow light in atomic vapors, the research groups of Ulf Leonhardt and Friedrich König collaborated to predict and measure the classical frequency shifting that is a signature of the Hawking radiation effect, also known as the positive-frequency Hawking radiation (PHR). This work was published in \textit{Science} in 2008 \cite{Philbin.2008}.

The experiment consists on sending a pump pulse through an optical fiber while simultaneously injecting a continuous wave (CW) probe to interact with the pump pulse by its induced change in refractive index due to the Kerr effect or SPM. The pump consisted of a 100 fs soliton with 800 nm central wavelength with a 76 MHz repetition rate. The probe is a CW laser of 1500 nm central wavelength. The fiber is a 1 m photonic-crystal fiber (PCF). The choice of the PCF is crucial to tailor the dispersion profile, ensuring a finite region of anomalous group-velocity dispersion (GVD) and enabling the horizon condition mentioned in Section \ref{sec.dispersion}.
		
\begin{figure}
	\centering
	\includegraphics[width=0.5\linewidth]{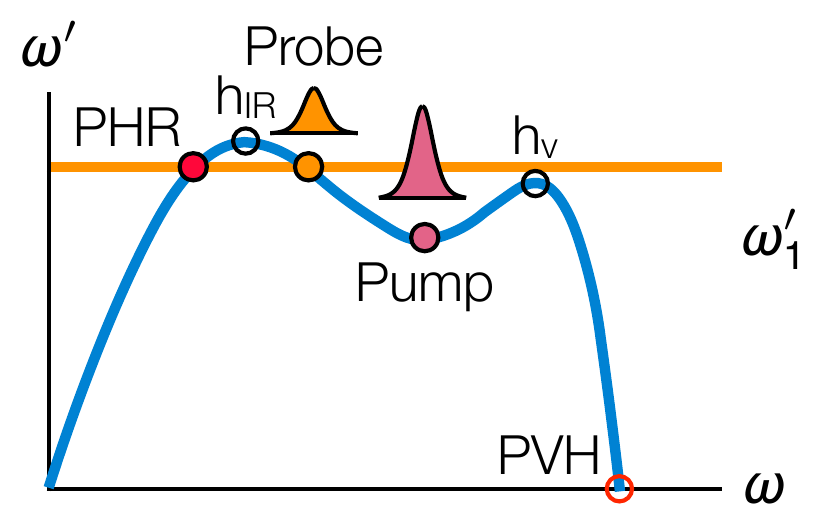}
	\caption{Dispersion relation illustrating the frequency shifting under the conservation of comoving frequency $\omega'$. PVH is the phase-velocity horizon, and h${}_\text{IR}$ and h${}_\text{v}$ are the infrared and visible group-velocity horizons.}
	\label{fig.PHR}
\end{figure}

As shown in Section \ref{sec.quantization}, the pump and the fiber constitute the background that creates the effective moving geometry where the probe is studied. In this configuration, the pump breaks the laboratory temporal and spatial symmetries. Consequently, neither laboratory energy nor momentum is conserved in this condition. However, a propagation time $\zeta=z/u$ symmetry can be recovered, such that the comoving frequency $\omega'$ is conserved $\Delta \omega'=0$.

The probe pulse propagates under this conservation law. In this condition, the comoving frequency is conserved, such that the prediction of the emitted frequency shifting or PHR is
\begin{equation}
	\omega'_\text{PHR}=\omega'_\text{probe}.
\end{equation}
This condition is illustrated in Fig. \ref{fig.PHR}. For convenience, we represent the shape of a generic dispersion relation with a finite anomalous GVD for clarity. The actual dispersion relation is qualitatively the same and leads to the same physical predictions; it is just slightly less convenient to plot a real dispersion relation (see Fig. 1 of reference \cite{Drori.2019}). A critical feature of this plot is the group-velocity horizon, defined as the frequency $\omega_h$ where the group velocity matches the pump velocity $v_g(\omega_\text{h})=v_g(\omega_\text{pump})=u$. This equality translates to
\begin{equation}
	\frac{\dd\omega'(\omega)}{\dd\omega}=0.
\end{equation}
For this shape of the dispersion relation,two frequencies fulfill this horizon condition $h_\text{ir}$ and $h_\text{v}$ for infrared and visible, respectively.
		
\begin{figure}
	\centering
	\includegraphics[width=0.5\linewidth]{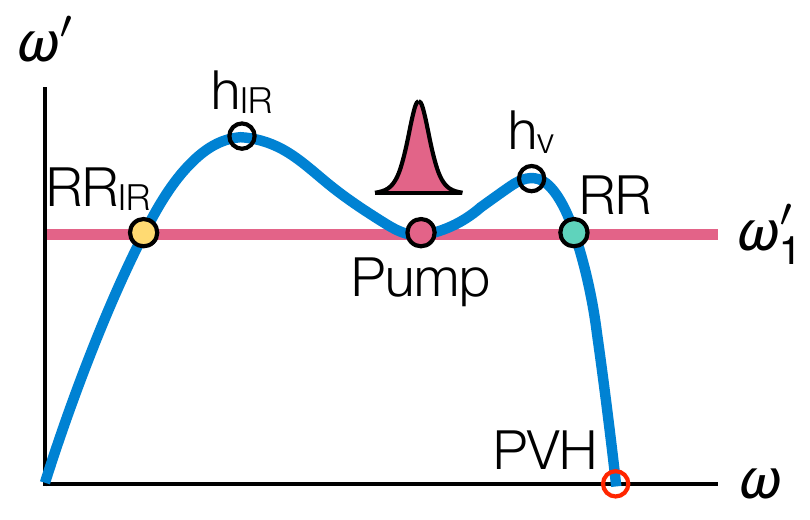}
	\caption{Dispersion relation $\omega'(\omega)$ illustrating that the conservation of $\omega'$ of the pump leads to the conditions for resonant radiation (RR).}
	\label{fig.RR}
\end{figure}

Under these conditions, the interaction between pump and probe in the comoving frame resembles the kinematics of a black-hole horizon at the leading edge of the pulse, inducing a redshifting of the probe frequency. Conversely, the trailing edge of the pulse acts like a white-hole horizon, producing a blueshift of the probe frequency, see Fig. \ref{fig.Kerr}. The model predicts the shifted laboratory frequencies solely through the comoving frequency conservation, as shown in Fig. \ref{fig.RR}.
		
It is also worth noting that this was the first time that the conservation of comoving frequency $\Delta \omega'=0$ was used to predict a new optical signal. Notably, fulfilling the so-called resonant condition, i.e., the conservation of comoving energy $\Delta \omega'=0$, is sufficient to create new output frequencies, bypassing the need to fulfill the phase-matching conditions. The phase-matching conditions include the laboratory energy conservation $\Delta \omega=0$  and laboratory momentum conservation $\Delta k=0$. The 2008 theoretical prediction and experimental detection seem to confirm this fact. However, an existing process can also be reinterpreted and explained under such resonant condition. It is the so-called Cherenkov radiation in optical fibers, also known as dispersive wave and resonant radiation. This is why we revisit a previous experiment in the next section.
		
\subsection{The 2003 Pavia experiment revisited}\label{sec.2003}		
The first predicted and measured effect fulfilling the resonant condition has been studied for decades; it is the resonant radiation, though it is also known by several different names, including dispersive wave, Cherenkov radiation, and non-solitonic radiation. The phenomenon was likely first predicted in optical fibers via numerical calculations by Wai et al. at the University of Maryland in 1987 \cite{Wai.1987}.

In 1995, the seminal paper by Akhmediev and Karlsson \cite{Akhmediev.1995} reinterpreted the dispersive wave emitted by a soliton in an optical fiber as a Cherenkov effect occurring in the (1+1)D comoving frame of a soliton or pump pulse propagating in an optical fiber. The emitted signal must fulfill the condition \cite{Rubino.2012}
\begin{equation}
	\beta(\omega_{\text{RR}})=\beta{\omega_{\text{pump}}}+\frac{\omega_{\text{RR}}-\omega_{\text{pump}}}{v_\text{g}}.
\end{equation}
This expression is mathematically equivalent to the conservation of comoving frequency
\begin{equation}
\Delta \omega'=0,
\end{equation}
based on our derivation in Section \ref{sec.NLSEcm}. Consequently, this is now also known as resonant radiation and the condition of conservation of comoving frequency is known as resonant condition. The intensity of the emitted signal is proportional to the spectral power of the soliton at the predicted resonant frequency. For this reason, generating resonant radiation requires ultra-short pulses---typically 100 fs or less---and high intensities--- usually with soliton number $N\simeq 2$. This regime, characterized by the breakdown of standard approximations, is known as extreme nonlinear optics (XNLO) \cite{Drori.2019}—the same domain required for observing the optical analogue of Hawking radiation.

The first experimental measurement had to wait until 2003 by Tartara et al. in Pavia \cite{Tartara.2003}. A crucial step was the development of photonic-crystal fibers (PCFs) by St. J. Russell et al. in 1996, see Ref. \cite{Russell.2003}. The experimental setup was the following. A pulsed laser with 190 fs duration at 810 nm with $N=1$, i.e., fundamental solitons, is coupled to a 40-cm PCF, and report a blue-shifted narrow visible component at 430 nm, the resonant radiation. This signal constituted the first experimental observation of resonant radiation, achieving a conversion efficiency of up to 24\%.

\subsection{The 2012 Heriot-Watt and St. Andrews experiments}\label{sec.2012}
Following the clarification of the physical reality of the resonant condition, the same group that work on the 2008 St. Andrews experiment realized that the prediction of the analogue Hawking effect carried a deeper implication: the existence of a Hawking partner possessing negative comoving frequency. At the time, it was controversial whether these negative comoving frequencies represented real physical states \cite{Biancalana.2012}  or if they could meaningfully mix with positive frequencies to create new detectable signals. However, the team hypothesized that the most accessible signature of this negative-frequency sector would not be the Hawking partner itself, but a negative-frequency resonant radiation (NRR). Mathematically, instead of having the conservation with positive comoving frequency as in RR $\omega'_\text{RR}=\omega'_\text{pump}$, NRR corresponds to the condition
\begin{equation}
	\omega'_\text{NRR}=-\omega'_\text{pump}.
\end{equation}
		
Standard dispersion models rarely accounted for the ultraviolet (UV) region where this signal was anticipated. However, using an extended dispersion model, the team predicted that NRR, if it existed, would manifest in the UV range, at around 230 nm. A collaboration between the experimental groups of Heriot--Watt University---led by Daniele Faccio---and of St. Andrews University---led by Friedrich König---supported by the theoretical group of Ulf Leonhardt, was formed to measure this signal. The resulting work presents two sets of measurements: one in optical filaments in bulk media, and the other in PCFs. These measurements provided a crucial step in demonstrating the physical reality of negative-frequency modes and their nonlinear coupling with positive-frequency fields.
		
\begin{figure}
	\centering
	\includegraphics[width=0.6\linewidth]{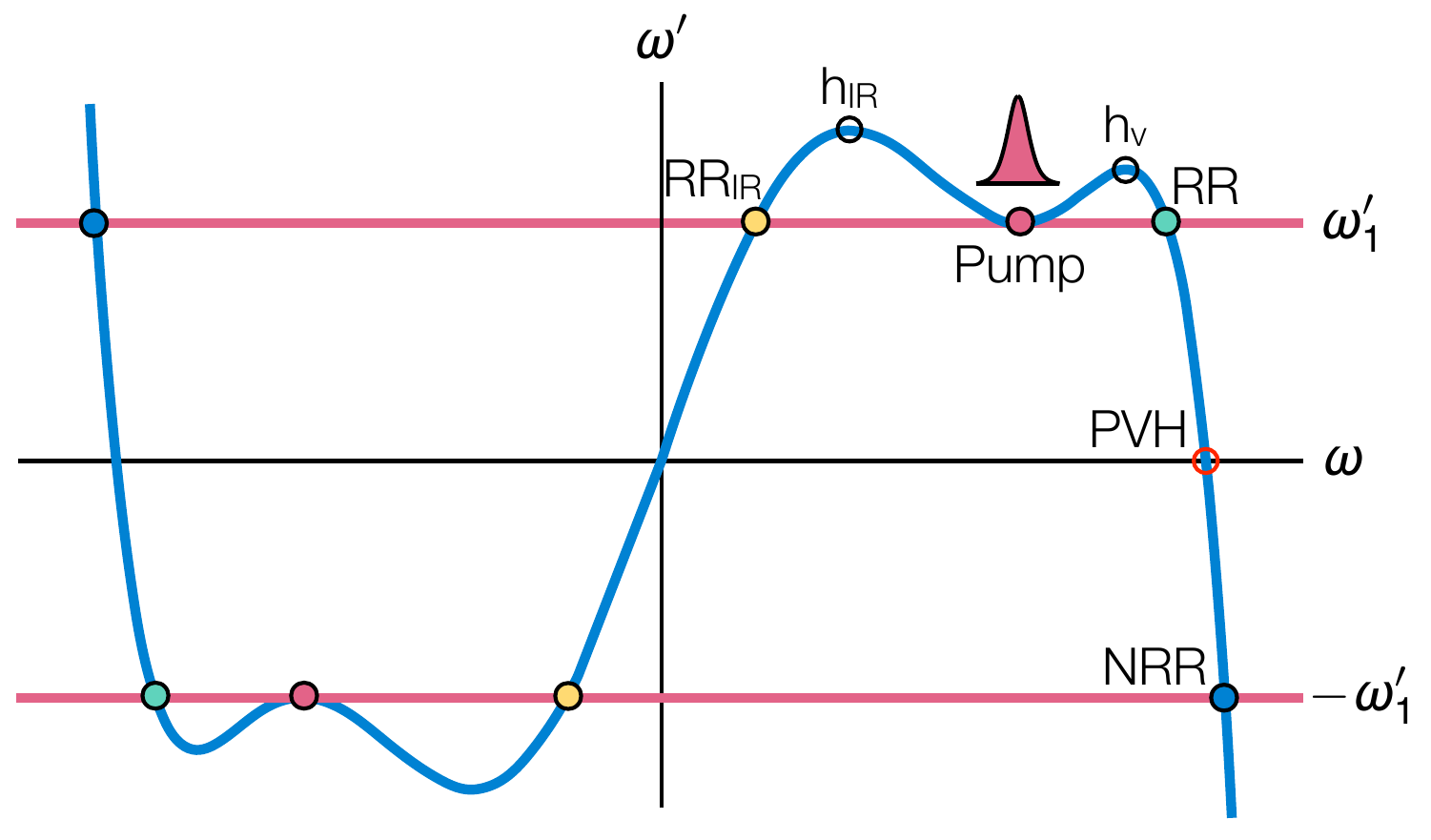}
	\caption{Dispersion relation $\omega'(\omega)$ showing the condition for the negative-frequency resonant radiation (NRR) at $-\omega_1'$.}
	\label{fig.NRR}
\end{figure}
	
The predicted NRR signal arises from the matching condition where the pump's phase-matching line intersects the negative-frequency branch of the dispersion relation in the comoving frame. In the 2012 experiments, the researchers employed various PCFs and successfully measured the NRR in the range of 218--233 nm. This discovery confirmed that the "negative frequency" side of the optical spectrum is not merely a mathematical artifact but a dynamical participant in extreme nonlinear optics.
		
\subsection{The 2012 St. Andrews experiment}\label{sec.2012b}
In 2012, the group of Friedrich König developed and tested a theory for the efficiency of the PHR based on the solution of a wave equation, taking the form of the NLSE and, under the pump-probe approximation, the actual Schrödinger equation \cite{Choudhary.2012}.
		
As established in Section \ref{sec.NLSElab}, pulse propagation is fundamentally modeled by the NLSE. However, within the pump-probe approximation, the equation for the probe field is linearized. This linearization reduces the system to a Schrödinger-type equation, where the pulse intensity $I(\tau)$ acts as an effective potential barrier and the comoving frequency $\omega'$—specifically its distance from the horizon frequency $\omega'_\text{h}$—serves as the effective energy of the particle.
		
\begin{figure}
	\centering
	\includegraphics[width=0.6\linewidth]{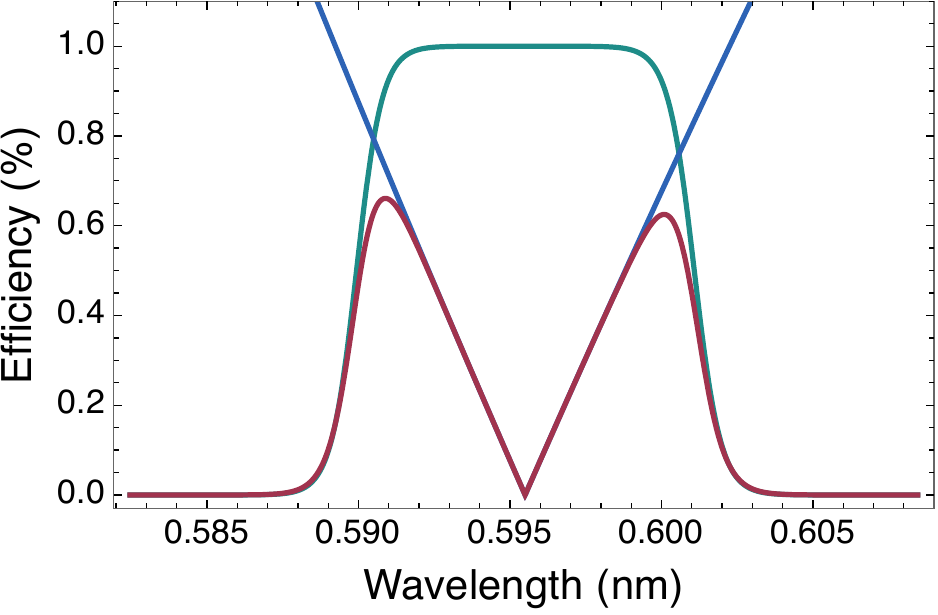}
	\caption{The efficiency of the frequency shifting around the visible horizon at 595 nm (red). It is the product of two factors: the amount of light colliding with the horizon (blue) and the reflection coefficient (green).}
	\label{fig.efficiency}
\end{figure}

The team conducted a series of experiments to validate this tunneling model. The efficiency of the stimulated Hawking emission depended critically on the group-velocity mismatch between the pump pulse and the CW probe. Specifically, the conversion efficiency is governed by how closely the probe frequency approaches the horizon frequency $\omega'_\text{h}$, where the potential barrier is most effectively "tunneled" or reflected. The resulting efficiency profile follows the characteristic curve shown in Fig. \ref{fig.efficiency}.

This experimental verification was significant because it moved beyond the mere observation of frequency shifts and began to characterize the scattering amplitudes of the analogue horizon, providing a quantitative link to the underlying QFTCS framework.

\subsection{The 2019 Weizmann experiment}\label{sec.2019}
The logical progression from the 2012 experiments was the direct measurement of the Hawking partner, also known as Negative-frequency Hawking Radiation (NHR). However, detecting this mode proved significantly more complex than anticipated. Although the NHR was predicted to appear in the spectral vicinity of the previously measured NRR, it was found that the conversion efficiency when using a CW probe was insufficient for detection.

\begin{figure}
			\centering
			\includegraphics[width=0.6\linewidth]{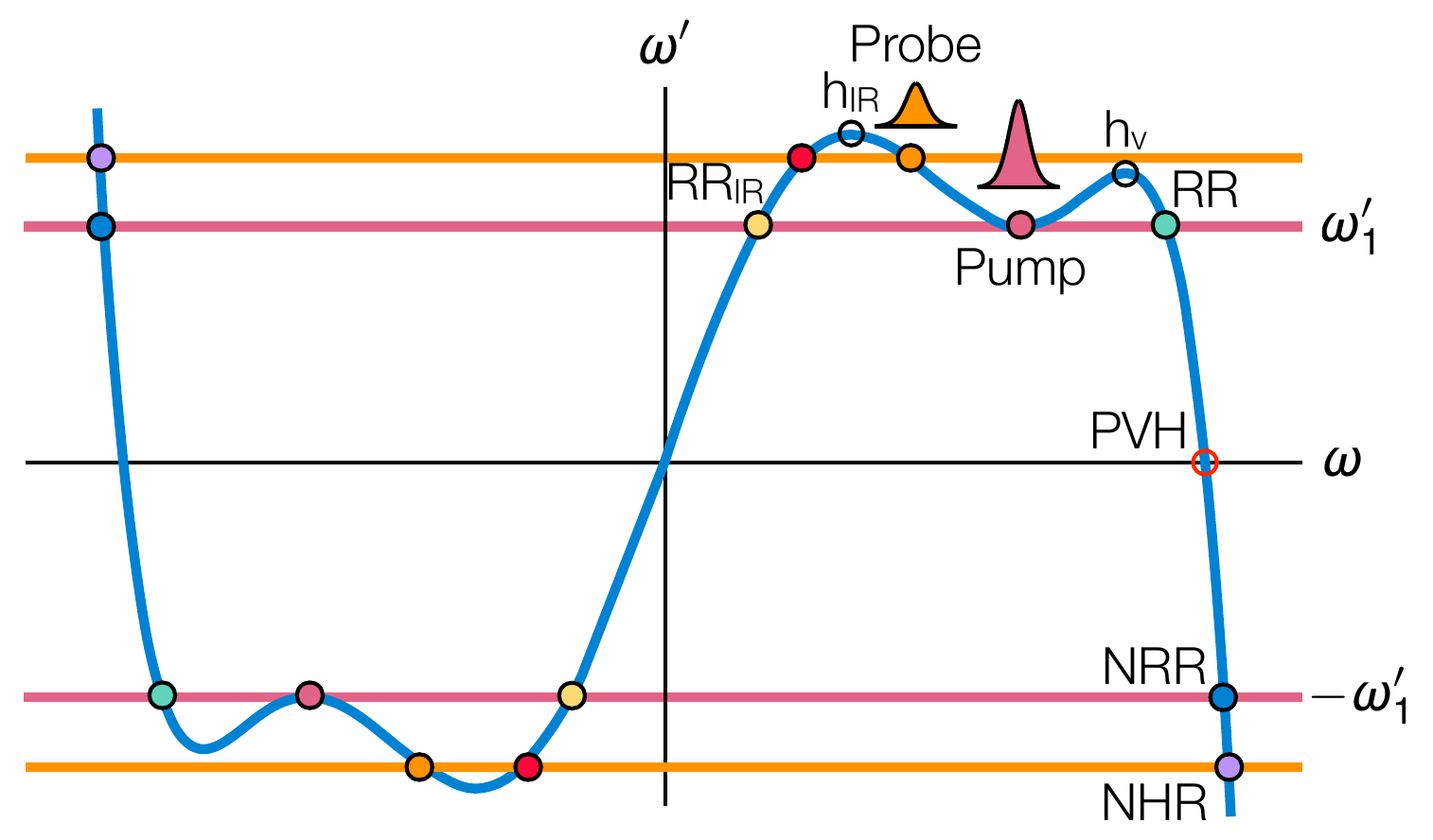}
			\caption{Dispersion relation $\omega'(\omega)$ showing the condition for the emission of negative-frequency Hawking radiation (NHR).}
			\label{fig.NHR}
\end{figure}

Consequently, a pulsed probe was required to enhance the nonlinear interaction, a task that demanded precise sub-picosecond synchronization between two ultra-short pulses. In 2019, Ulf Leonhardt's group at the Weizmann Institute of Science successfully reported the measurement of NHR. To ensure temporal coherence, both the pump and the probe were derived from the same master laser. Specifically, the probe was tuned to the infrared horizon through Raman intrapulse frequency shift (RIFS) \cite{Rosenberg.2020}. The initial pulse was divided; one part was sent through a pre-propagation fiber where soliton self-frequency shifting redshifted it toward the horizon frequency (as shown in Fig. \ref{fig.NHR}).

After this preparation, the pump and the redshifted probe were coupled into an interaction fiber, where they were synchronized to overlap temporally and generate the NHR. The NHR follows a resonant condition similar to the PHR, but with a negative comoving frequency:
\begin{equation}
	\omega'_\text{NHR}=-\omega'_\text{probe}.
\end{equation}
Notably, this condition produces a signal that is blueshifted by only approximately 1 nm relative to the NRR signal. To isolate the NHR from this nearby background, two distinct experiments were performed: one utilizing only the pump to record the NRR baseline in the UV, and a second using both the pump and probe to capture the combined NRR and NHR spectrum. Subtracting the baseline from the combined measurement provided the first clean evidence of the stimulated Hawking partner. The experiment used 8 fs pulses at 800 nm in 7 mm PCF.
		
\subsection{The 2022 UNAM experiment}\label{sec.2022}
In 2022, a research group at the National Autonomous University of Mexico (UNAM) reported the first measurement of the interference between two Positive-frequency Hawking Radiation (PHR) signals. By launching two synchronized probe pulses to interact with a single pump soliton, they demonstrated that the resulting Hawking emissions were phase-locked. This interference experiment confirmed the coherence property of the Hawking signal—a critical requirement for future quantum correlation studies.

Notably, the group utilized constructive interference to achieve a fourfold increase ($4\times$) in the peak intensity of the signal without distorting its spectral shape. This experiment was also significant because it was carried out at the second group-velocity horizon located in the visible spectrum near 650 nm (see Fig. \ref{fig.NHR}). By successfully shifting the interaction to this visible frequency range, the group provided further evidence for the robustness of the resonant condition in predicting measured signals in the laboratory across different dispersion regimes.
		
\begin{figure}
	\centering
	\includegraphics[width=0.6\linewidth]{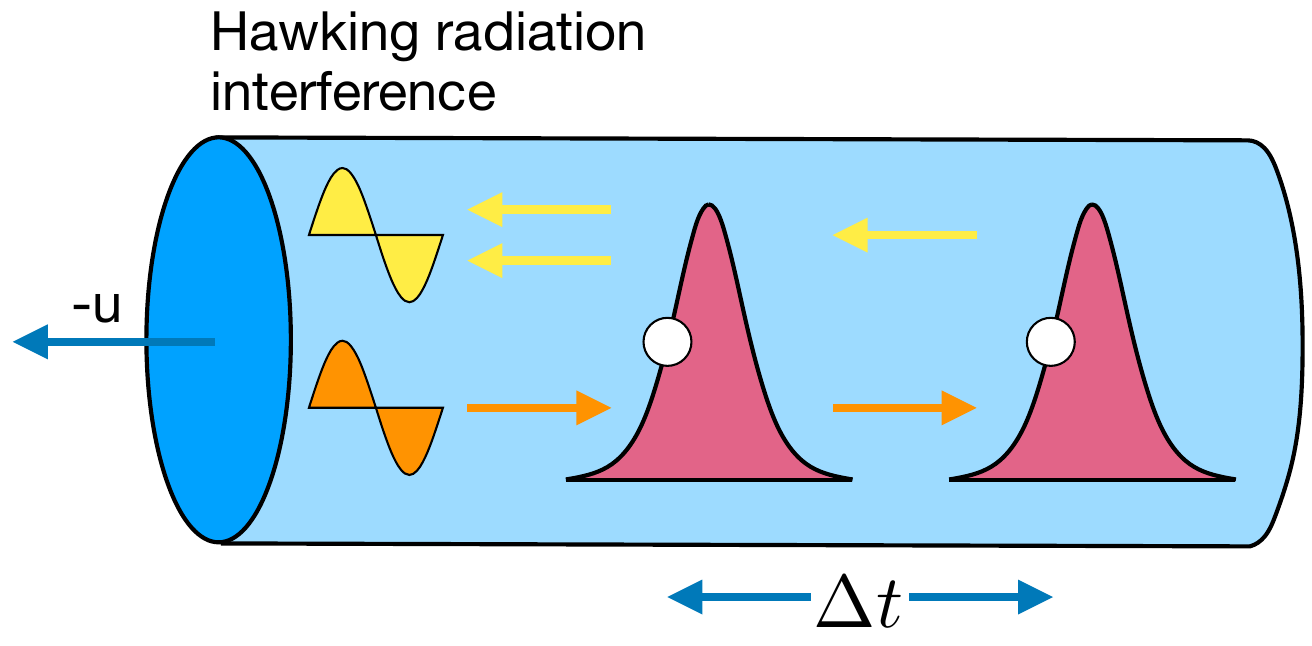}
	\caption{Interference of two analogue Hawking radiation signals ()yellow) stimulated by a CW probe (orange) interacting with a double pump pulse (red).}
	\label{fig.inter}
\end{figure}
		
\subsection{Current experiments}\label{sec.now}
This concludes our short overview of the milestone experiments related to the optical analogue of Hawking radiation. However, significant challenges remain. In particular, nearly all experiments discussed so far operate in the stimulated, classical regime rather than the spontaneous, quantum regime. Active research is currently focused on reaching the quantum regime, alongside efforts to expand the theory of resonant signals by moving beyond the standard approximations used in extreme nonlinear optics (XNLO).

The field remains remarkably active, and new experimental breakthroughs are expected shortly. A primary objective is the direct measurement of spontaneous Hawking radiation in optics, which would be highly welcomed by the community. The fiber-optic platform is particularly well-suited for this task because the direct detection of single photons is a common occurrence in modern optics labs. Furthermore, several statistical tests of "quantumness" can be performed between the Hawking radiation and its partner to verify their predicted quantum entanglement, providing a definitive signature of the Hawking effect.
	
\section{Conclusions}\label{sec.conclusions}
During these notes, we have discussed the theoretical and experimental landscape of analogue gravity in optical fibers. This field bridges the gap between black hole physics and nonlinear quantum optics. The core of this work rests on the formal analogy between light in a moving medium and a field in a curved spacetime. As we established, the refractive index gradient of a laser pulse acts as a coordinate transformation, where the group velocity of the pulse creates a blocking horizon for probe waves or for the vacuum state of a quantum probe field. This mapping allows us to treat the kinematic properties of a black hole---such as the event horizon and surface gravity---within the experimental environment of a fiber-optics laboratory.

By leveraging the unidirectional pulse propagation equation (UPPE) and the nonlinear Schrödinger equation (NLSE), we demonstrated how an intense soliton traveling through a dispersive medium creates this moving geometry. Our derivation of the Lagrangian and Hamiltonian formalisms and the subsequent application of Noether's theorem highlighted the fundamental role of symmetries in these systems. We demonstrated that while laboratory energy and momentum are not conserved due to the pump's presence, the comoving frequency $\omega'$ remains invariant. This conservation law is the fundamental resonance condition that predicts the frequency shifts of positive-frequency Hawking radiation (PHR), its negative-frequency partner (NHR), and the closely related resonant radiation (RR) and its negative-frequency equivalent (NRR).

Finally, we decided to discuss the key experiments in an historical timeline---from the first proposal at St. Andrews to the recent breakthroughs at Weizmann and UNAM. The evolution of the experimental results underscores a remarkable transition in the field. We have moved from the mere observation of frequency shifts in 2008 to the sophisticated detection of the Hawking partner in 2019 and the verification of signal coherence in 2022. These milestones prove that the "negative-frequency" side of the optical spectrum is a dynamical reality, accessible through the tools of extreme nonlinear optics (XNLO).

As we look toward the future, the ultimate goal remains the unambiguous detection of spontaneous quantum Hawking radiation. While the experiments discussed herein primarily utilize stimulated emission, the platform of photonic-crystal fibers (PCFs) is uniquely positioned to enter the quantum regime. The high precision of single-photon counting and the ability to perform entanglement verification between the Hawking mode and its partner offer a path toward observing "true" Hawking pairs.

In conclusion, the study of optical horizons does more than just simulate black holes; it validates the robustness of quantum field theory in curved spacetime (QFTCS) and challenges our understanding of light-matter interactions at the extreme. The fiber-optical analogue stands as a testament to the universality of physics, proving that the secrets of the cosmos can indeed be found within a meter (or less) of glass.

\section*{Acknowledgments}
The authors would like to thank Ivan Agullo, Maxime Jacquet, Dimitrios Kranas, Nicholas Pavloff, Scott Robertson, Justin Wilson, and Iacopo Carusotto for valuable discussions. 

\paragraph{Funding information}
DB would like to thank the support of the Marcos Moshinsky Chair (2024) and of the Secretariat of Science, Humanities, Technology and Innovation (Mexico) SNI-62547.





\bibliography{Library2026.bib}


\end{document}